\documentclass[11pt]{article}

\usepackage[utf8]{inputenc}
\usepackage{jheppub}
\usepackage[export]{adjustbox}
\usepackage[english]{babel}
\usepackage{amsmath}
\usepackage{color}
\usepackage{epstopdf}
\usepackage{graphicx}
\usepackage{braket}
\usepackage{axodraw4j}
\usepackage{amsmath}
\usepackage{amsmath}
\usepackage{slashed}
\usepackage{amssymb}
\usepackage{braket}
\usepackage{amsmath}
\usepackage{braket}
\usepackage{epigraph}
\usepackage{tensor}
\usepackage{enumerate}
\usepackage{subfig}
\usepackage{verbatim}
\usepackage{hyperref}

\usepackage{mathtools}
\usepackage{tikz}
\usetikzlibrary{matrix,arrows,decorations.pathmorphing,decorations.pathreplacing}

\usepackage{tikz}
\usetikzlibrary{matrix,arrows,decorations.pathmorphing,decorations.pathreplacing}
\usepackage{amsmath,amssymb,euscript,array,mathrsfs,appendix,ctable,marvosym}
\usepackage{arydshln}
\usepackage{todonotes}
\usepackage{tikz}
\usepackage{pgfplots}
\pgfplotsset{/pgf/number format/use comma,compat=newest}
\usepackage{color}
\usepackage{graphicx}  

\author{B. Garbrecht, J. I. McDonald}

\affiliation{Technische Universität München,\newline Physik-Department, James-Franck-Straße, \newline
85748 Garching, \newline
Germany}

\emailAdd{garbrecht@tum.de}
\emailAdd{jamie.mcdonald@tum.de}

\title{Axion configurations around pulsars}

\abstract{Axion-like particles lead to a plethora of new phenomena relating to compact astrophysical objects including stellar and black hole superradiance, axion stars and axion clusters.  In this work, we investigate a new scenario in which macroscopic axion configurations are sourced by the electromagnetic fields of pulsars via the axion-photon coupling.  We solve the inhomogeneous axion field equation with an explicit source term given by the electromagnetic fields of the pulsar in a rotating magnetic dipole approximation. We find that the axion profile either forms a localised boundstate or radiates as outgoing waves, depending on whether the pulsar frequency is smaller or greater than the axion mass, respectively. We derive the total mass of the scalar cloud generated around the pulsar, and the power loss for radiative solutions.  We point out that it will be necessary  to incorporate pulsar magnetosphere effects and their partial screening of the axion-photon interaction  for a more accurate quantitative prediction. Finally, we suggest some observational signatures which should be investigated in future work. }

\begin{document}

\maketitle

\section{Introduction}

Axion physics currently represents one of the most active areas at the interface of particle theory, astrophysics and cosmology. Motivated by the axion solution to the strong CP-problem \cite{Peccei:1977hh,Peccei:2006as,Marsh:2015xka} and axion-like particles arising from string theory \cite{Svrcek:2006yi,Arvanitaki:2009fg} we consider a generic scalar field which interacts with electromagnetism (in  Lorentz–Heaviside units $\mu_0 = \varepsilon_0=1$) via the following Lagrangian minimally coupled to gravity:
\begin{equation}
\mathcal{L} = \sqrt{-g}\left[\frac{1}{2}\partial_\mu \phi \partial^\mu \phi - \frac{1}{2}\phi^2 \mathcal{M}^2_\phi - \frac{1}{4} F_{\mu \nu}F^{\mu \nu} - \frac{g_{\gamma \phi}}{4}  \phi F_{\mu \nu}\tilde{F}^{\mu \nu} \right], \label{AxionL}
\end{equation}
where $\tilde{F}_{\mu \nu} = \varepsilon_{\rho \sigma \mu \nu}/2F^{\rho \sigma}$. Using $F_{\mu \nu} \tilde{F}^{\mu \nu} = - 4 \textbf{E}\cdot \textbf{B}$, the axion equation of motion becomes 
 \begin{align}
 \square \phi  + \mathcal{M}_\phi^2 \phi = g_{\gamma \phi}\textbf{E}\cdot \textbf{B}. \label{AxionEOM}
 \end{align}

 This paper seeks to make the first steps towards understanding a simple question: what is the nature of scalar clouds sourced by an $\textbf{E}\cdot\textbf{B}$ component of the strong electromagnetic fields of pulsars, and what might its observational signatures be?
 
 Constraining the strength of the axion-photon coupling $g_{\gamma \phi}$ is the focus of many current and future experiments, e.g.~haloscopes \cite{Sikivie:1983ip} such as ADMX \cite{Asztalos:2009yp}, and MADMAX \cite{TheMADMAXWorkingGroup:2016hpc,Millar:2016cjp,Majorovits:2016yvk} helioscopes CAST \cite{Anastassopoulos:2017ftl}, IAXO \cite{Irastorza:2011gs}, and light-shining through a wall tests like ALPS-II \cite{Dobrich:2013mja} (see \cite{Redondo:2010dp,Adler:2008gk} for review), to name but a few. 
 
 In comparison to many other astrophysical sources, the electromagnetic fields of pulsars \cite{Hewish:1968bj} can reach some of the strongest known values in the Universe, with field strengths of up to $10^{15}\text{G}$ in the case of magnetars \cite{Harding:2013ij}. This makes pulsars an excellent playground in which to study the axion-photon coupling, whose strong magnetic fields can compensate for the large supression of $g_{\gamma \phi}$. It should be noted that we consider macroscopic axion field configurations, rather than individual scattering processes in the cores of stars \cite{Raffelt:1990yz} used to place stellar cooling bounds on axion parameters. Thus our solutions are more in the vein of other classical axion configurations around astrophysical objects which we now discuss.

There already exists a wide array of novel astrophysical predications should the axion turn out to exist in nature. These are of merit not only as tools to further constrain the physics of ultralight scalar degrees of freedom but also as astrophysical phenomena in their own right. Perhaps the most widely studied effect involving axions and compact astrophysical objects is superradiance, in which scalar fields extract rotational energy from a spinning black hole \cite{Arvanitaki:2010sy,Pani:2012vp,Arvanitaki:2014wva,Brito:2014wla,Brito:2015oca,Rosa:2017ury} or star \cite{Cardoso:2015zqa,Richartz:2013hza,Cardoso:2017kgn} via an instability for massive perturbations.  Other notable signatures relate to pulsar timing, in which pressure oscillations from a background axion field distort spacetime, causing an apparent frequency change in pulses \cite{Khmelnitsky:2013lxt}. Similarly axion stars \cite{Liebling:2012fv,Braaten:2015eeu,Visinelli:2017ooc,Helfer:2016ljl} present  an entirely new class of astrophysical objects whose detection has recently been considered in \cite{Bai:2017feq,Sigl:2017sew}. Clusters  \cite{Enander:2017ogx} also represent yet another kind of axion-based astrophysical object whose radio signatures have been examined in \cite{Tkachev:2014dpa}. It is remarkable to think that such a rich array of phenomena arise from something as simple as a scalar field and it is in this spirit that we explore axion clouds around pulsars. 
\\\\
 \noindent \textbf{Contrasting with superradiance}\newline
It is worth remarking that the nature of axion solutions explored in this paper is qualitatively different from discussions of superradiance in the context of black holes and stars mentioned in the preceding paragraph. The superradiant scenario essentially describes a \textit{homogeneous} wave equation for a particle ``trapped in a box" which exhibits a fundamental set of eigenfrequencies $\omega = \omega_R + i \omega_I$ with an imaginary part $\omega_I$ whose sign is such that solutions grow exponentially in time. Crucially, the scalar field does not corotate and has $\omega \neq \Omega$, where $\Omega$ is the frequency of the spinning object. In fact the superradiant constraint is more stringent, requiring $\omega_R< m \Omega$, where $m$ is the azimuthal quantum number of the mode in question. The physical picture underlying this mechanism is that rotating objects can dissipate energy onto an incoming wave, leaving the scattered wave with more energy so that $|A_\text{out}|/|A_\text{in}|>1$ where $A_\text{out}$ and $A_{\text{in}}$ are the amplitudes of the outgoing and incoming waves, respectively. If one repeats this process ad infinitum by placing a ``mirror" near the black hole or star, the incoming wave comes back with  more energy each time. The analogue of such a mirror is to give the field a non-vanishing mass, which has the effect of confining it to a region near the black hole.
 
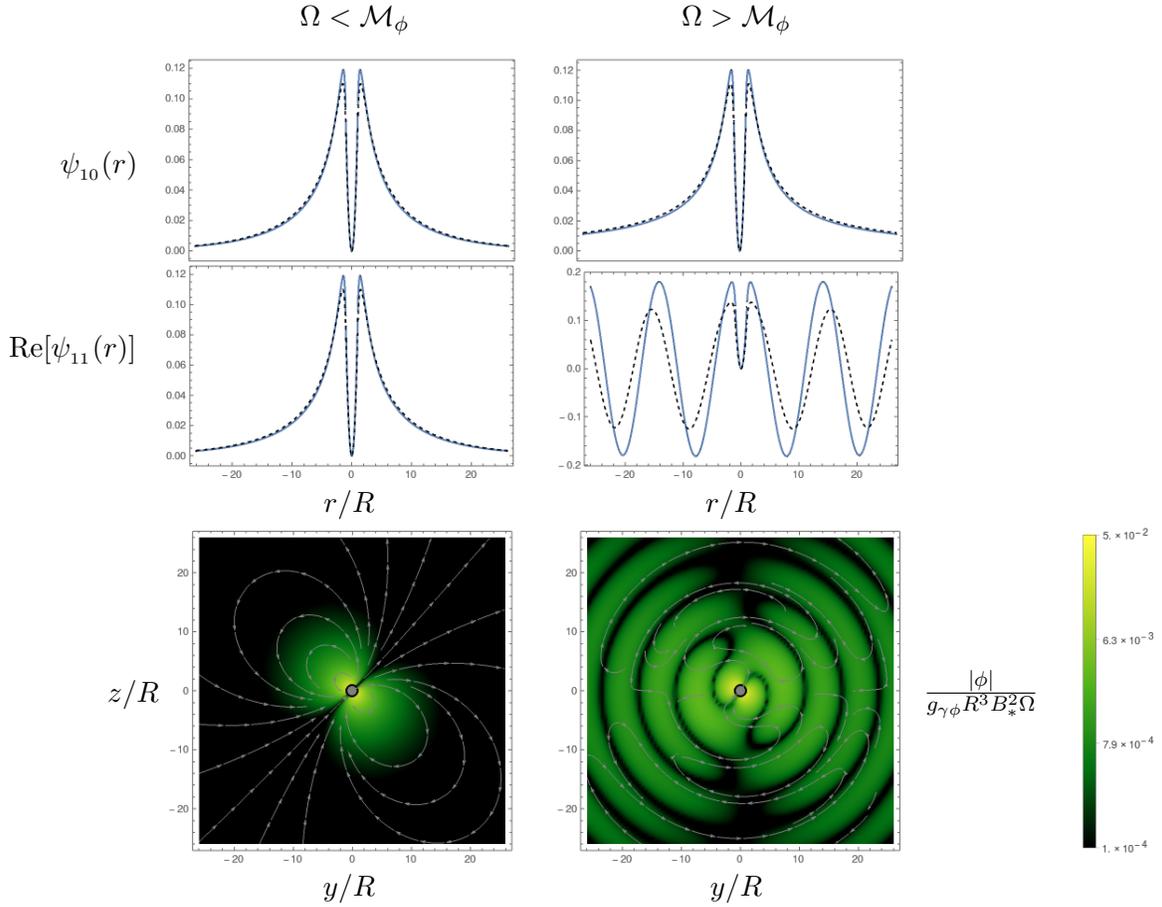
\begin{figure}[ht]
  	\pgfdeclareimage[interpolate=true]{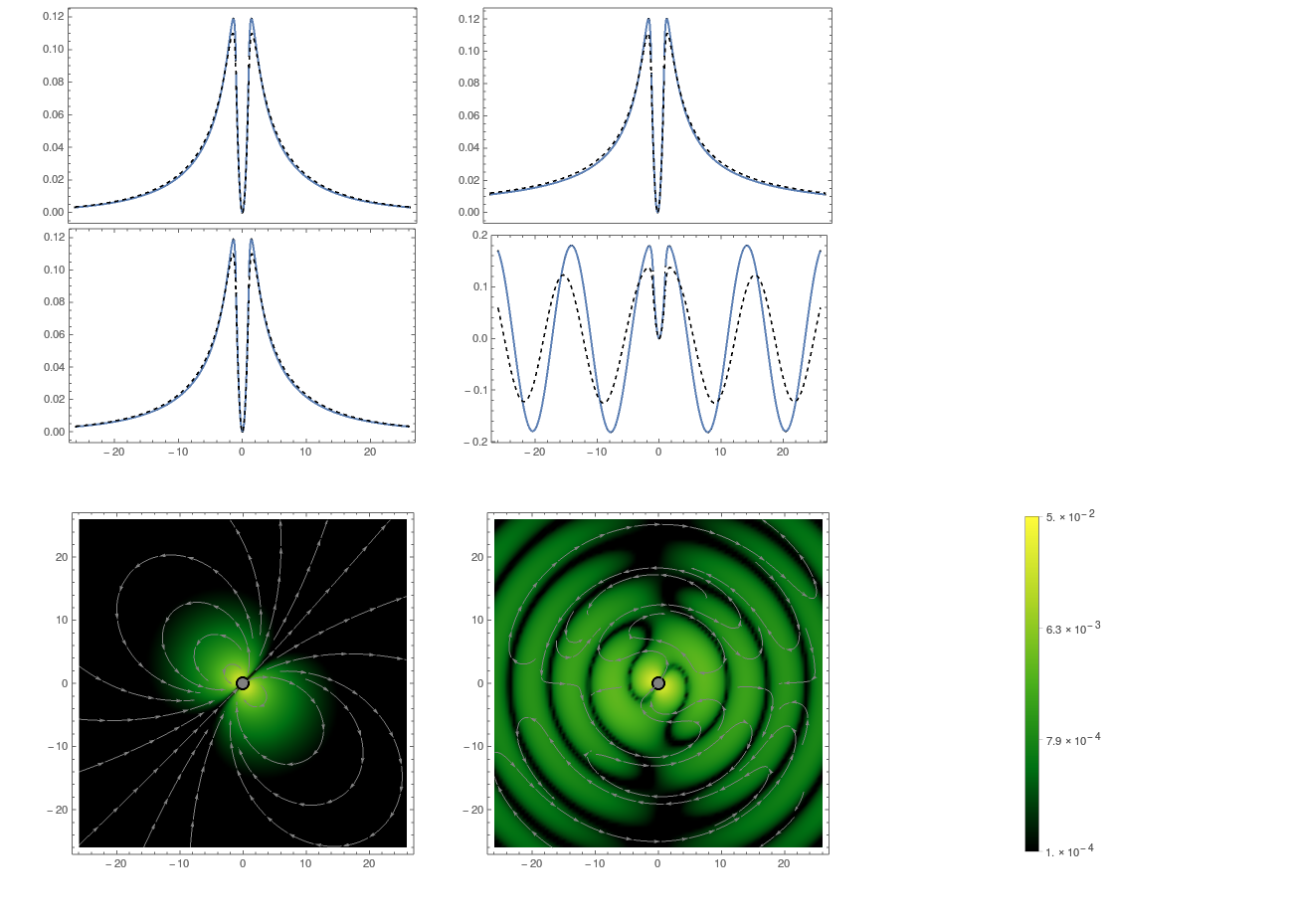}{ContourPlots.png}
  	\begin{center}
  		\begin{tikzpicture}[scale=0.35]
  		\pgftext[at=\pgfpoint{0cm}{0cm},left,base]{\pgfuseimage{ContourPlots.png}}
  		\node at (8.5,1) {$y/R$};
  		\node at (23.2,1) {$y/R$};
  		\node at (8.5,15.6) {$r/R$};
  		\node at (23,15.6) {$r/R$};
  		\node at (0.3,8.4) {$z/R$};
  		\node at (-2,21.5) {$\text{Re}[\psi_{_{11}}(r)]$};
  		\node at (-1,28.5) {$\psi_{_{10}}(r)$};
  		\node at (32.5,8.4) {$\frac{\left|\phi\right|}{g_{\gamma \phi}R^3 B_*^2  \Omega}$};
  		\node at (8.7,34) {$\Omega <\mathcal{M}_\phi$};
  		\node at (23.2,34) {$\Omega >\mathcal{M}_\phi$};
  		\end{tikzpicture}
  	\end{center}
  	\caption{Axion contours sourced by the electromagnetic fields of the pulsar in the vacuum dipole model (VDM), as seen in the  $\textbf{M}$ - $\boldsymbol{\Omega}$ plane. We took $\mathcal{M}_\phi R = 0.1$ and $r_s/R=0.3$. The left column shows a boundstate with $ \Omega R = 0.01$ whilst the right shows radiating solutions with $\Omega R=0.5$. The upper two rows show the axion radial function (solid) together with its analytic approximation (dashed) given by neglecting gravity and far field corrections to EM fields described in sec.~\ref{Analytic}. In the bottom plots, the magnetic field lines are shown in grey.  }
  	\label{AerialWave}
  \end{figure}

  By contrast, in our setup, the scalar field continually extracts energy from the $\textbf{E}\cdot \textbf{B}$ source term in the \textit{inhomogeneous} wave equation (\ref{AxionEOM}). This generates a configuration which, under certain circumstances (specifically when $\mathcal{M}_\phi < \Omega$), consists of a purely \textit{outgoing} wave ($|A_\text{in}|=0$) at infinity as the axion solution radiates away energy from the system. Furthermore, by linearity, the scalar field contribution sourced by the electromagnetic fields inherits the same frequency as the pulsar so that by contrast with the superradiant case, we have instead a corotating solution $\omega = \Omega$.  From a calculational point of view then, we must go further than simply solving a homogeneous wave equation with particular boundary conditions. We shall actually derive Green functions and integrate them with a source to construct axion profiles. It is also worth noting that the scenario considered here has the advantage that it probes the axion-photon coupling directly. 
 \\\\
  \noindent \textbf{Summary of results}
  \begin{itemize}
  	\item We compute axion solutions to the inhomogeneous wave equation (\ref{AxionEOM}) by a method of Green functions, with the source given by the electromagnetic fields of a pulsar in an oblique rotating magnetic dipole approximation.  Solutions are shown in fig.~\ref{AerialWave}.
  	\item The maximum density of solutions grows with decreasing axion mass. For $\mathcal{M}_\phi > \Omega$, solutions are localised near the pulsar with an amplitude falling off exponentially at large distances. For $\mathcal{M}_\phi < \Omega$ they radiate as outgoing axion waves. For very low masses $\mathcal{M}_\phi \ll \Omega$, densities tend to a maximum value (fig.~\ref{Peak}) with the  largest axion densities (see eq.~(\ref{rhoAxion})) reaching  $\sim 10^{-1} \text{g}\, \text{cm}^{-3}$ for typical neutron star and axion parameters. We find that in the extremal case, for the largest known pulsar magnetic fields of $B_* \sim 10^{15} \text{G}$ and the upper bound for the axion photon copuling $g_{\gamma \phi} = 10^{-13} \text{GeV}^{-1}$, the total mass of the cloud can reach $\sim 10^{-13} M_\odot$, a comparable size to certain axion stars and clusters. 
  	\item In the radiative regime $\mathcal{M}_\phi < \Omega$, we compare the axion power output to the magnitude of electromagnetic dipole radiation - see eq.~(\ref{axionpower}) and fig.~\ref{Power}.
  \end{itemize}
  \vspace{0.5cm}

  The structure of this paper is as follows. In the following two sections we introduce our basic model, beginning with the gravitational metric in sec.~\ref{gravMetric}. We then outline the structure of electromagnetic fields around the pulsar in sec.~\ref{PulsarB} in a simple rotating magnetic dipole model. There we explain how the axion field satisfies an inhomogeneous wave equation with a source term derived from the dot product of the pulsar's electric and magnetic fields. In sec.~\ref{RadialSec} we describe axion solutions to this wave equation, deriving the radial component of solutions and offering an approximate analytic solution in sec.~\ref{Analytic}. We then compute the total mass of the cloud in sec.~\ref{MassSec} which for reference we compare to the typical masses of axion stars. We also describe the total power radiated by the axion cloud in sec.~\ref{PowerSec} and compare this to the power of electromagnetic dipole radiation. In sec.~\ref{screening} we suggest how our stellar  modelling could be improved to incorporate plasma around the pulsar and how this will localise the axion cloud to regions in which $\textbf{E}\cdot \textbf{B}$ is poorly screened by the pulsar magnetosphere. Finally in sec.~\ref{discsusion} we offer our conclusions and give some possible observational effects which should be pursued in future work.

\section{Gravitational Metric}\label{gravMetric}

In our choice of gravitational metric we shall neglect rotational effects (i.e.~frame dragging etc.) that would be encountered in e.g.~a Kerr metric \cite{1963PhRvL..11..237K}. This has the benefit of greatly simplifying the form of the wave equation of the scalar field. We shall therefore consider a Schwarzschild geometry matched to an interior solution corresponding to uniform stellar density. The Schwarzschild metric is a good approximation provided the ratio of the star's total angular momentum to mass $J/M$, is much smaller than its radius $R$, in which case Kerr gives the limiting case of the Schwarzschild metric. As a rough approximation, treating the pulsar as a sphere of uniform density one has $J =\frac{2}{5} \Omega M R^2$,  so that $J/M = \frac{2}{5} R (\Omega R) $. Note that  $\Omega R$  is precisely the tangential speed of the pulsar at the equator and therefore must be less than the speed of light.  For typical pulsars, $R \sim 10\text{km}$ and $\Omega \lesssim 10^{2} \text{Hz}$ so that $\Omega R \ll 1$. In this case, $J/M \ll R$ making the Schwarzschild metric an excellent approximation for most pulsars.

Explicitly then, our metric is of a Schwarzschild form matched to a perfect fluid of uniform density in the stellar interior \cite{griffiths_podolsky_2009,Schwarzschild:1916ae} so that the full metric takes the form
\begin{equation}
ds^2 =  e^{2 \Psi} dt^2 - N^{-1}dr^2 + r^2 \left(d\theta^2 + \sin^2 \theta d \varphi^2  \right),
\label{SchwarzsInt}
\end{equation}
where 
\begin{align}
N(r) & = 
\left \{
\begin{tabular}{cc}
$1- \frac{ r_s r^2}{R^3}$ &  $\qquad \quad r\leq R$\\
$1 - \frac{r_s}{r}$ &   $\qquad  \quad r > R$
\end{tabular} 
\right. , \nonumber \\ 
&\nonumber \\
e^{\Psi} & =  
\left \{
\begin{tabular}{cc}
$\frac{3}{2}\sqrt{1 - \frac{r_s }{R}} - \frac{1}{2}\sqrt{1 - \frac{ r_sr^2}{R^3}}$ &  $\qquad \quad r\leq R$\\
$\sqrt{1 - \frac{r_s}{r}}$ &   $ \qquad  \quad r > R$
\end{tabular}
\right. , \label{metric}
\end{align}
where $r_s= 2GM$ is the Schwarzschild radius of the star. Using the relation
\begin{equation}
\nabla_\mu \nabla^\mu \phi = \frac{1}{\sqrt{-g}} \partial_{\mu} \left(\sqrt{-g} g^{\mu \nu} \partial_{\nu} \phi \right),
\end{equation}
it is easy to see the axion wave equation then satisfies 
\begin{align}
&e^{-2 \Psi} \frac{\partial^2 \phi}{\partial t^2} - \frac{1}{ r^2 N^{-1/2} e^{\Psi}} \frac{\partial}{\partial r} \left( r^2 N^{1/2} e^{\Psi} \frac{\partial \phi}{\partial r} \right) - \frac{1}{r^2 \sin \theta}\frac{\partial}{\partial \theta}\left( \sin \theta \frac{\partial \phi}{\partial \theta}\right) 
- \frac{1}{r^2 \sin^2 \theta} \frac{\partial^2 \phi}{\partial \varphi^2} + \mathcal{M}_\phi^2 \phi \nonumber \\
&= g_{\gamma \phi}\textbf{E}\cdot\textbf{B},
\label{SchwarzsWave}
\end{align}
where $(r,\theta, \varphi)$ are spherical polar coordinates. Consider a monochromatic source for the wave equation of the form
\begin{equation}
g_{\gamma \phi}\textbf{E}\cdot\textbf{B} =\frac{ e^{-i \Omega t}}{r} \sum_{\ell, m} j_{\ell m }(r) Y_{\ell m}(\theta, \varphi), \label{YDecomp}
\end{equation}
where $Y_{\ell m}(\theta, \varphi)$ are spherical harmonics. In accordance with eq.~(\ref{SchwarzsWave}) we shall consider axion solutions of the form
\begin{equation}
\phi = \frac{e^{-i\Omega t} }{r}  \sum_{\ell, m} \psi_{\ell m}(r) Y_{\ell m}(\theta, \varphi) . \label{axionAnsatz}
\end{equation}
Of course, since the wave equation (\ref{SchwarzsWave}) is linear, a monochromatic source will only excite axion modes of the same frequency and quantum number as the source, so that, up to the addition of a homogeneous solution arising from some other unspecified physics, the axion cloud should have the same time dependence as the source.  However, in a regime where axion self-interactions are important, one could have further axion modes excited. One could envisage a situation in which the source first fills some quantum numbers, which then begin to fill other harmonics at different frequencies due to axion self-interactions, as happens in superradiance and bosanova effects \cite{Yoshino:2012kn,Yoshino:2015nsa}. For instance, in the case of a cosine potential $U(\phi) = \mathcal{M}_\phi^2 f_a^2 \left[1 - \cos(\phi/f_a) \right]$ non-linear effects would become important if the axion cloud reached densities such that $\phi \sim f_a$. However, for the sake of progress we shall consider the simplest scenario in which we focus only on a linear setup with corotating axion solutions of the form (\ref{axionAnsatz}), whose substitution into (\ref{SchwarzsWave}) gives the following radial wave equation for the $\psi_{\ell m}(r)$:

\begin{equation}
- A \frac{d }{d r} \left(A \frac{d \psi_{\ell m}}{\partial r} \right) + 
\left[
e^{2\Psi}\left(\frac{l(l+1)}{r^2} + \mathcal{M}_\phi^2\right) + \frac{A' A}{r}
\right] \psi_{\ell m} = \Omega^2 \psi_{\ell m} +  j_{\ell m}(r) e^{2\Psi},
\end{equation}
where primes denote derivatives with respect to $r$ and 
\begin{equation}
\qquad A  = e^{\Psi} N^{1/2}. 
\end{equation}
This can be thought of as a Sch{\"o}dinger equation with an extra source term with effective potential given by
\begin{equation}
V_{\ell}(r) = 
e^{2\Psi}\left(\frac{l(l+1)}{r^2} + \mathcal{M}_\phi^2\right) + \frac{A' A}{r}.
\label{Vl}
\end{equation}

The homogeneous part of the radial equation has been well-studied in the Schwarzschild case \cite{Chandrasekhar:1975zza,Jensen:1985in,Leaver:1986gd, Nollert:1992ifk,Konoplya:2004wg}  we consider here and also in its extension to Kerr backgrounds \cite{Brill:1972xj,1977RSPSA.352..381D,Kokkotas:1999bd,Hod:2012px}, the main interest being in superradiance and quansinormal modes. There are many approaches to solving the homogeneous equation \cite{Konoplya:2011qq}, ranging from series solutions to numerical shooting methods. In general though, with the exception of a few special cases, no simple analytic solutions are known and one must resort to some form of numerics to get the full solutions.  

At first glance, one might also worry that by neglecting Kerr effects, one is missing important instabilities of the scalar field arising from its extraction of rotational energy from the star - so called \textit{superradiance} \cite{Brito:2015oca} encountered in the context of rotating black holes. However, this mechanism arises from the special form of boundary conditions which are imposed on the wave equation at the black hole horizon. It is this boundary condition which gives rise to a quantised set of eigenmodes exhibiting an imaginary frequency part which leads to instabilities which grow exponentially in time.

 Since stars do not possess a horizon, the wave equation in its simplest form does not give rise to superradiant effects of this nature. However, it is worth noting that if there is some source of dissipation for the scalar field then superradiance can occur for stars. For instance, if a friction term term $\alpha \dot{\phi}$ is added to the left hand side of (\ref{AxionEOM}) (where $\alpha$ is some dissipation constant), then superradiant solutions are possible, as shown in \cite{Cardoso:2015zqa}. In principle, such a dissipative term could arise from quantum corrections to the classical equations of motion. For instance, it could be generated by the imaginary part of the one-loop photon correction to the axion propagator, which by the optical theorem corresponds to $\phi \rightarrow \gamma \gamma$ decays, in analogy with the mechanism described in ref.~\cite{Yokoyama:2005dv}.   
 Similarly, for dark photons around a star, subject to some version of Ohm's law, a finite conductivity provides a mechanism for dissipation \cite{Richartz:2013hza,Cardoso:2017kgn}, in which case dark photon solutions exhibit superradiance provided they have a non-vanishing mass.

In the present case, we shall not explore these possibilities and consider the simplest setup, consisting only of a wave equation with a source $\textbf{E}\cdot \textbf{B}$. Our calculation will proceed as follows. For the exterior solution, we shall generate the axion profile by first constructing a Green function for the radial equation in the region $r\geq R$, which we then integrate with the source terms to get the full exterior solutions, subject to appropriate boundary conditions at infinity. In these exterior solutions it remains to fix a boundary condition at the stellar surface, which we do by imposing continuity of the solution at $r=R$. For black hole spacetimes, one imposes a stringent boundary condition by demanding that modes at the horizon should be purely ingoing, so that the horizon acts as a one-way membrane.  This gives rise to a set of discrete eigenfrequencies - the quasinormal modes of the black hole \cite{Kokkotas:1999bd}.  Since stars do not possess a horizon, the imposition of such boundary conditions at the stellar surface would be artificial. Instead, we should demand only that the solutions satisfy appropriate continuity conditions at the stellar surface and match smoothly onto an interior solution.

\section{Pulsar electromagnetic field structure}\label{PulsarB}

Next we must choose a model for the electromagnetic fields inside and outside the pulsar. In order to solve the axion equation of motion, we shall make two simplifying assumptions. Firstly we assume that the interior of the star is a perfect conductor, in the sense that the electric field $\textbf{E}'$ in the rest frame of the star vanishes:
\begin{equation}
\textbf{E}' = 0,  \qquad \qquad 0 \leq r \leq R.
\end{equation}
Using the transformation laws for electromagnetic fields in rotating frames \cite{1970AmJPh..38.1487M, 1984IJTP...23..441G}, the fields $\textbf{E}$ and $\textbf{B}$ as seen in the inertial frame are related to $\textbf{E}'$ by
\begin{equation}
\textbf{E}'=\textbf{E} + \textbf{v} \times \textbf{B} = 0, \qquad \textbf{v} = \boldsymbol{\Omega} \times \textbf{r}, \qquad \text{for}  \qquad 0\leq  r \leq R, \label{PerfectConductor}
\end{equation}
where $\textbf{r}$ is the position vector with respect to an origin at the centre of the star. Eq.~(\ref{PerfectConductor}) then implies that the source of the axion wave equation (\ref{SchwarzsWave}) vanishes in the stellar interior:
\begin{equation}
\textbf{E}\cdot \textbf{B} = 0, \qquad 0 \leq r \leq R.
\end{equation}
 Of course in reality, the axion experiences in medium effects due to the nuclear matter inside the star which could modify its potential, as has recently been examined in \cite{Hook:2017psm}. This could change the profile of the interior axion solution. However, for the sake of making progress we leave such considerations for future work.

Next we discuss the structure of the electromagnetic fields outside the star. We shall choose perhaps the simplest, and one of the earliest models of pulsar magnetic fields. This is the classic vacuum dipole model (VDM) of Deutsch \cite{1955AnAp...18....1D} which, in accordance with the assumption above of a perfectly conducting stellar interior, models the pulsar by a rotating dipole, inclined by an angle $\alpha$ relative to the axis of rotation. In this setup, one neglects any plasma surrounding the pulsar, so that the external electromagnetic fields are in vacuo. The effects of plasma surrounding a pulsar and its implications for $\textbf{E}\cdot \textbf{B}$ are discussed briefly in sec.~\ref{screening}.

 Working in Coulomb gauge with $\nabla \cdot \textbf{A}=0$ and $A^0 = 0$, the vacuum equations for the gauge potential become
\begin{equation}\label{AEOM}
\frac{\partial^2 \textbf{A}}{\partial t^2} -  \nabla^2 \textbf{A} =0,
\end{equation}
with the electric and magnetic fields are given by
\begin{equation}
\textbf{E} = -\frac{\partial \textbf{A}}{\partial t}, \qquad \textbf{B} = \nabla \times \textbf{A}.
\end{equation}
The VDM is characterised by the rotating dipole equations
\begin{equation}
\textbf{A} = \frac{1}{4 \pi} \nabla \times \left[ \frac{\textbf{M}(t-r)}{r} \right], \qquad \frac{\partial \textbf{M}}{\partial t} = \boldsymbol{\Omega} \times \textbf{M}, \label{GaugeA}
\end{equation}
where $\textbf{M}(t)$ is the rotating, inclined dipole moment.  Explicitly one has (see e.g. \cite{Melrose:2011eh})
\begin{align}
\textbf{E} & = \frac{1}{4 \pi}\left[\frac{\textbf{r} \times \dot{\textbf{M}}}{r^3} + \frac{\textbf{r} \times \ddot{\textbf{M}}}{r^2}\right] ,\label{Efield} \\
& \nonumber\\
\textbf{B}& =  \frac{1}{4 \pi}\left[
\frac{ 3 \textbf{r}(\textbf{r}\cdot \textbf{M}) - r^2\textbf{M}}{r^5}
+
\frac{ 3 \textbf{r}(\textbf{r}\cdot \dot{\textbf{M}}) - r^2 \dot{\textbf{M}}}{r^4} 
+
\frac{\textbf{r} \times (  \textbf{r} \times \ddot{\textbf{M}})}{r^3}
\right] \label{BfielD}.
\end{align} 
Notice that since
\begin{equation}
\dot{\textbf{M}} = \boldsymbol{\Omega} \times \textbf{M}, \qquad  \ddot{\textbf{M}} = \boldsymbol{\Omega} \times (\boldsymbol{\Omega} \times \textbf{M}),
\end{equation}
it follows that in the ``near zone", characterised by $r \Omega\ll 1 $, the second term in $(\ref{Efield})$ and the last two terms in the expression for $(\ref{BfielD})$ are sub-leading, giving the following near-zone approximations 
\begin{equation}
\textbf{B}_{\text{near}} =\frac{B_* R^3}{2r^3}\left(3 (\hat{\textbf{M}} \cdot \hat{\textbf{r}}) \hat{\textbf{r}} - \hat{\textbf{M}} \right), \qquad 
\textbf{E}_\text{near} = \frac{B_* R^3 \Omega}{2r^2} \hat{\textbf{r}} \times( \hat{\boldsymbol{\Omega}} \times \hat{\textbf{M}}), \label{near}
\end{equation}
where hats denote unit vectors and
\begin{equation}
B_* = \frac{2  M}{4 \pi R^3},
\end{equation}
is the value of the magnetic field at the dipole on the stellar surface. The solutions (\ref{near}) can also be derived by taking retarded time $t-r \rightarrow t$ in (\ref{GaugeA}), so that they correspond to the neglect of retarded effects and electromagnetic radiation. 

One can find an explicit expression for $\textbf{M}(t)$ by solving the second constraint in (\ref{GaugeA}). This can be achieved by taking a dipole aligned with $\boldsymbol{\Omega}$ (which we choose as the $z$-axis) and performing two successive SO$(3)$ rotations, the second of which depends on time. One first takes a vector of constant magnitude $M$, pointing in the $z$-direction and rotates it by an angle $\alpha$ relative to the $z$ axis - this makes the dipole oblique. Then one sets it spinning by performing a rotation about the $z$-axis through an angle $\Omega t$. This can be written as
\begin{equation}
\textbf{M}(t)  = 
\left(
\begin{array}{ccc}
\cos \Omega t   & -\sin \Omega t & 0   \\
\sin \Omega t   &  \cos \Omega t & 0   \\
0               &   0            & 1
\end{array}
\right)
\left(
\begin{array}{ccc}
1  &     0      &  0            \\
0  & \cos \alpha  &  \sin \alpha   \\
0  & -\sin \alpha  &  \cos \alpha
\end{array}
\right)
\left(
\begin{array}{c}
0 \\
0 \\
M
\end{array}
\right)= 
M \left(
\begin{array}{c}
-\sin \alpha  \sin t \Omega   \\
 \sin \alpha  \cos t \Omega  \\
\cos \alpha 
\end{array}
\right). \label{M}\\
\end{equation}
Inserting the expression (\ref{M}) into (\ref{Efield}) and (\ref{BfielD}), one finds, after a lengthy but straightforward calculation that
\begin{equation}
\textbf{E}\cdot \textbf{B} = \frac{R^6}{r^5}\frac{B_*^2 \Omega }{4}   \sin \alpha \Big( \cos \theta  \sin \alpha +\sin \theta  \cos \alpha  \left[ r \Omega  \cos (\lambda +r \Omega )-\sin (\lambda
+r \Omega ) \right]\Big),\quad  R \leq r \leq \infty, \label{E.B}
\end{equation}
where in (\ref{E.B}) we have introduced the coordinate
\begin{equation}
\lambda  = \varphi  - \Omega t,
\end{equation}
which we recognise as the azimuthal coordinate in the rest frame of the star. There is therefore no explicit $t$ dependence other than that contained in $\lambda$, from which we see that the electromagnetic fields corotate with the pulsar. Notice that the terms  $\mathcal{O}(r \Omega)$ in the arguments of the trigonometric functions correspond to electromagnetic radiation, giving outgoing spherical waves $\sim e^{i \Omega(r-t)}$, such effects are therefore neglected in the near field limit. Note also that $\textbf{E}\cdot\textbf{B}$ consists of a stationary axisymmetric part and time-dependent part with non-trivial azimuthal dependence.

We notice immediately that the expression (\ref{E.B}) can be rewritten in terms of the harmonics
\begin{equation}
Y_{10}(\theta, \varphi) = \frac{1}{2}\sqrt{\frac{3}{\pi}}\cdot \cos \theta, \qquad \quad  Y_{11}(\theta, \varphi) = -\frac{1}{2}\sqrt{\frac{3}{2\pi}} e^{i \varphi}\cdot \sin \theta,
\end{equation}
so that
\begin{align}
&\textbf{E}\cdot \textbf{B} = 
\left\{
\begin{array}{cc}
\frac{R^6 B_*^2 \Omega}{2 r ^5}\sqrt{\frac{\pi }{3}} \text{Re} 
\left[ 
\sin^2 \alpha Y_{10}(\lambda,\theta)
- \frac{\sin 2 \alpha}{\sqrt{2}} (1 -  i r \Omega)\,e^{i r \Omega}  \,i \, \text{Y}_{11}(\lambda,\theta)
\right] & \quad R < r < \infty  \\
0 & \quad0 \leq r\leq R
\end{array}
\right. ,
 \label{EBSOURCE}
\end{align}
where we have absorbed the explicit $e^{-i \Omega t}$ dependence into the spherical harmonics via $\lambda$. The profile of $\textbf{E}\cdot \textbf{B}$ is shown in figure \ref{EdotBPlot}. 
\pgfdeclareimage[interpolate=true]{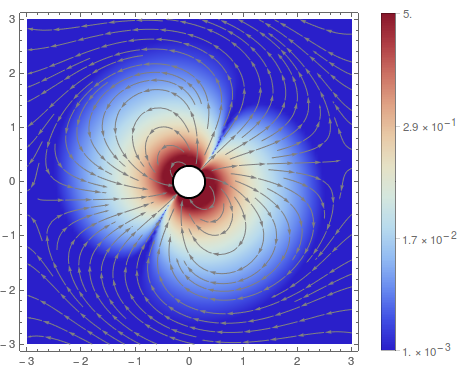}{EdotBPlot1.png}
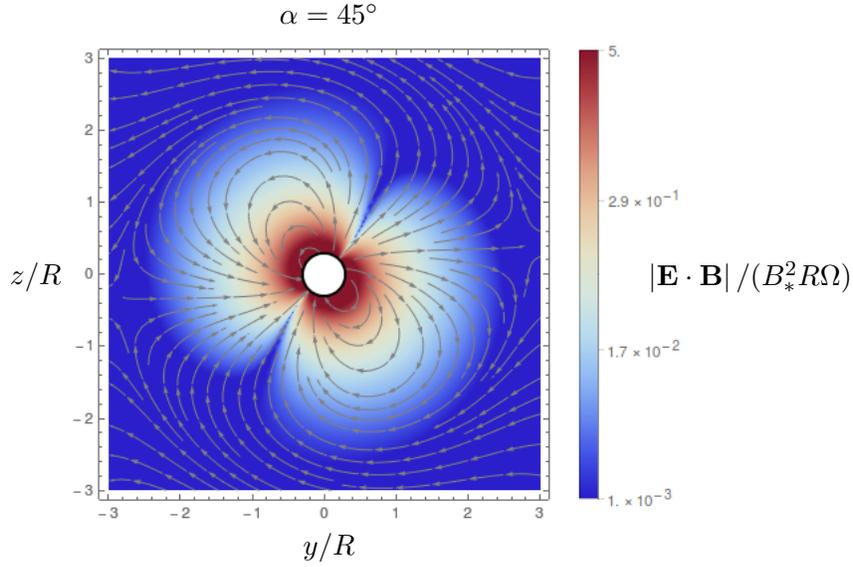
\begin{figure}[ht]
	\begin{center}
		\begin{tikzpicture}[scale=0.5]
		\pgftext[at=\pgfpoint{0cm}{0cm},left,base]{\pgfuseimage{EdotBPlot1.png}}
		\node at (6.8,-0.2) {$y/R$};
		\node at (-1,7) {$z/R$};
		\node at (18,7) {$\left|\textbf{E}\cdot\textbf{B}\right|/(B_*^2 R \Omega)$ };
		\node at (6.8,14) {$\alpha = 45 ^{\circ}$};
		\end{tikzpicture}
	\end{center}
	\caption{The absolute value of $\textbf{E}\cdot\textbf{B}$ (in units of $B_*^2 R \Omega$) in the $\textbf{M}$ - $\boldsymbol{\Omega}$ plane for the VDM with the value $ R \Omega=0.1$. The magnetic field lies are also shown (grey arrows). }
	\label{EdotBPlot}
\end{figure}

\section{Axion Solutions - Solving the radial equation}\label{RadialSec}

In accordance with our discussion in sec.~\ref{gravMetric}, a source of the form (\ref{EBSOURCE}) corresponds to an axion solution 

\begin{equation}
\phi(r, \theta, \lambda) =  \frac{1}{r} \frac{g_{\gamma \phi}R^4 B_*^2  \Omega}{2} \sqrt{\frac{\pi}{3}}  
\text{Re} 
\left[ 
\sin^2 \alpha \, \psi_{10}(r) Y_{10}(\lambda,\theta)
- \frac{\sin 2 \alpha}{\sqrt{2}} \psi_{11}(r)i\text{Y}_{11}(\lambda,\theta)
\right].
\label{PhiAnsatz}
\end{equation}
\\
\noindent The prefactor depending on axion and stellar parameters is inserted to simplify the final form of the radial equation. It also means that the radial functions $\psi_{\ell m}$ are dimensionless. Explicitly one finds the radial equations (with $\ell=1$)
\begin{align}
 \left[-A(r) \frac{d}{d r} \left( 
A(r)\frac{d }{d r}
\right) + V_\ell(r) -m^2 \Omega^2 \right]\psi_{\ell m} = 
\theta(r-R) \frac{R^2e^{2\Psi}}{r^4}\left\{ 
\begin{array}{cc}
 1 &  \quad m=0\\
 (1 - i r \Omega)e^{i r \Omega}  & \quad   m=1
\end{array}
\right., \label{radial1psi}
\end{align}
where $V_\ell$ is the effective potential given by eq.~(\ref{Vl}) and $\theta(r -R)$ is the heaviside step function. The $m=0$  mode corresponds to a stationary axissymmetric solution, whilst $m=1$ gives a co-rotating configuration which violates axisymmetry via a non-trivial $\varphi$ dependence. Notice that the $m=0$ mode therefore depends only on the axion mass (and not $\Omega$) whilst for $m=1$, the radial function depends depends both on $\Omega$ and $\mathcal{M}_\phi$. Notice that the radial equation (\ref{radial1psi}) can be written purely in terms of dimensionless variables by taking $(r,\mathcal{M}, \Omega) \rightarrow (r/R,\,\mathcal{M} R,\,\Omega R)$, so that the shape of the axion profile is determined purely by the two dimensionless parameters $\Omega R$ and $\mathcal{M} R$.

\subsection{Exterior axion solution}
The exterior wave equation is inhomogeneous and can be solved by method of Green functions satisfying 
\begin{align}
\left[-A(r) \frac{d}{d r} \left( 
A(r)\frac{d  }{d r}
\right) + V_\ell(r)  -  m^2 \Omega^2 \right]  G(r,r') = \delta(r,r').\label{Grstar}
\end{align}
We shall construct the Green function in the standard way \cite{Leaver:1986gd} from two linearly independent mode functions $f$ and $g$ which satisfy the homogeneous version of (\ref{Grstar}). This leads to a Green function of the form (see appendix \ref{Green}) 
\begin{equation}
G(r_,r') =   \frac{\gamma}{A(r')}  f(r) f(r')+ \frac{1}{2k A(r')}\left[
\theta(r - r')g(r') f(r)  + 
\theta(r' - r)g(r) f(r') \right], \quad 
r,r' \geq R, \label{G}
\end{equation}
where the constant $\gamma$ is fixed by demanding continuity of $\psi$ and $\psi'$ at stellar surface. The mode functions $f$ and $g$ are chosen to have decaying and growing boundary conditions at infinity with the asymptotic behaviour
	\begin{equation}
	f  \rightarrow e^{-k r  } \left(1 +\frac{1}{r k}\right), \quad \qquad
	g  \rightarrow e^{r  k } \left( 1 - \frac{1}{k r} \right), \qquad 
	 \text{as} \qquad 
	 r \rightarrow \infty \label{Asymptotics}.
	\end{equation}
	We have also defined
	\begin{equation}
	k = \sqrt{\mathcal{M}_\phi^2 - m \Omega^2}, \qquad  \text{Re}(k) >0,\quad \text{Im}(k)<0, \, \qquad \text{with} \qquad  m = 0,\,1.\label{k}
	\end{equation}
	 Notice that the $m=1$ mode has distinct behaviour depending on relative size of $\Omega$ and $\mathcal{M}_\phi$. For $\mathcal{M}>\Omega$, the quantity $k$ is real and positive, whilst for $\mathcal{M} < \Omega$ we see $k$ is pure imaginary. The boundary conditions (\ref{Asymptotics}) and the sign choice in (\ref{k}) therefore ensures that the full solutions given by (\ref{G}) either decay or give purely outgoing waves at infinity in the cases $\Omega < \mathcal{M}_\phi$  and $\Omega > \mathcal{M}_\phi$, respectively.  In the latter case, $|k|$ should be thought of as the 3-momentum of the outgoing waves at infinity, whilst for the $\Omega < \mathcal{M}_\phi$ case it sets the scale over which solutions decay. Thus as expected, axion radiation can be produced by an $m=1$ mode.

 To compute the mode functions $f$ and $g$ we solve the homogeneous version of (\ref{Grstar}) and impose the boundary conditions (\ref{Asymptotics}) at sufficiently large $r = r_{\text{max}}$. We perform a consistency check of these numerical solutions by checking their Wronskian matches the analytic result $W =2k/A(r)$. We also test the mode functions against their flat space limit $r_s=0$ for which there are exact analytic expressions given in sec.~\ref{Analytic}.

 \subsection{Interior axion solution}
The interior solution satisfies a homogeneous wave equation 
\begin{equation}
-A(r) \frac{d}{d r} \left( 
A(r)\frac{d \psi }{d r}
\right) + V_\ell(r) \psi -  m^2 \Omega^2 \psi =0, \qquad  0 \leq r \leq R, \label{uEqInterior}
\end{equation}
 where the metric components take the interior form given by eq.~(\ref{metric}). Since $r=0$ is a regular singular point, one can use Frobenius' method (see Appendix \ref{InteriorSol}) to write solutions in the form 
 \begin{align}
  \psi_\text{int}(r) = R^{-s} r^s \tilde{\psi}(r) ,
  \end{align}
  where the prefactor $R^{-s}$ ensures both $\psi$ and $\tilde{\psi}$ are dimensioneless. $\tilde{\psi}$ then satisfies
  \begin{equation}
  r^2 \tilde{\psi}'' + r \left( P + 2 s\right) \tilde{\psi}'+ \left[ s(s-1) + s P + Q \right]\tilde{\psi} =0, \qquad \quad \tilde{\psi}(0) = 0, \label{utildeEQ}
  \end{equation}
  with
\begin{align}
P(r) &= \frac{r A'}{A}, \qquad Q(r) = \frac{1}{A^2(r)}\left[r^2 m^2 \Omega^2  
- e^{2\Psi}\left(l(l+1) + \mathcal{M}_\phi^2r^2\right) - r A' A  \right], \nonumber \\
\nonumber \\
s & = \frac{1}{2}\left[ 1 + \sqrt{1 + 4 \ell (\ell+1)} \right].  \label{PQeq}
\end{align}
 We have obtained the interior solution by solving equation (\ref{utildeEQ}) numerically with appropriate boundary conditions and arbitrary overall normalisation by choosing $\tilde{\psi}(0)=1$. The remaining  normalisation of $\psi_\text{int}$, is fixed by imposing continuity at the stellar surface, as we now describe.

\subsection{Full solution: matching at the stellar surface }
The full solution is then given by 
\begin{equation}
\psi_{\ell m}(r ) =  \left\{ 
\begin{array}{cc}
 f(r)\left[\alpha + \int^{r}_R dr' \frac{g\left(r'\right)}{2k A(r')}j \left(r'\right) \right]
+g(r)\int^{\infty}_r dr' \frac{f\left(r'\right)}{2k A(r')}j \left(r'\right)  & \qquad \infty > r \geq R  \\
 \beta  \psi_\text{int}(r)   &\qquad 0 \leq r \leq R
\end{array}
\right. , \label{solutionform}
\end{equation}
where $\beta$ gives the overall normalisation of the solution in the stellar interior and $\alpha = \gamma \int_R^\infty dr f(r)j(r)/A(r)$. The source $j$ corresponds exactly to the right hand side of eq.~(\ref{radial1psi}). The constant $\alpha$ and $\beta$ are uniquely determined by continuity of $\psi$ and its first derivative at the stellar surface. This leads immediately to
\begin{equation}
\alpha  = \left\{  
I \frac{ W\left[g,f\right]}{ W\left[ \psi_\text{int},f \right]} \frac{\psi_\text{int}}{f }
- \frac{g}{f}I 
 \right\}_{r =R},  \qquad \qquad \beta  = \left\{ I \frac{W\left[g,f\right]}{W\left[ \psi_\text{int},f \right]} \right\}_{r =R}, \label{AlphaBeta}
\end{equation}
where 
\begin{equation}
I (r) =  \int^{\infty}_r dr' \frac{f\left(r'\right)}{2k A(r') } j(r'). \label{I}
\end{equation}
With the equations (\ref{solutionform}) and (\ref{AlphaBeta}), we have now fully determined the solution for the axion field both inside and outside the pulsar in the VDM, together with the assumption of an interior Schwarzschild metric corresponding to a stellar interior of constant mass density. 

The steps involved in computing the full solution thus go as follows: for each value of $\mathcal{M}$ and $\mathcal{\omega}$, compute the homogeneous mode functions $f$ and $g$ (in our case numerically) and compute the interior homogeneous solution via $\tilde{\psi}_-$. Then insert $f$ and $g$ into the integrals with the source in (\ref{solutionform}) as well as equation (\ref{I}). Then one can use these to determine the matching coefficients via (\ref{AlphaBeta}). This should then be repeated for each mode $(\ell, m)$ after which one will have completely determined the axion solution for a given $\Omega$ and $\mathcal{M}_\phi$.

Notice that in the radiative regime ($\Omega> \mathcal{M}_\phi$) the solution at infinity consists of purely outgoing waves, unlike in superradiance which involves continual reflection off the central object. The solution takes the form
\begin{equation}
\psi = \mathcal{A} e^{i|k| r}  \qquad r \rightarrow \infty \label{uInfinity},
\end{equation}
with
\begin{align}
\mathcal{A} &=\alpha + \int^{\infty}_R dr' \frac{g\left(r'\right)}{2k A(r')}j \left(r'\right) \label{AmplitudeEq}.
\end{align}
Notice that the amplitude of the outgoing wave is essentially determined by what is going on near the stellar surface, since every quantity except the integral is evaluated at $r=R$ and even that is dominated by this region, since the source $j$ decays as $1/r^4$. 

%\begin{figure}[ht]
%	\pgfdeclareimage[interpolate=true]{AmpPlot.pdf}{AmpPlot.pdf}
%	\begin{center}
%		\begin{tikzpicture}[scale=0.4]
%		\pgftext[at=\pgfpoint{0cm}{0cm},left,base]{\pgfuseimage{AmpPlot.pdf}}
%		%		\node[rotate=-90] at (19.4,10.8) {$\Omega = \mathcal{M}_\phi$} ;
%		\node at (13.5,-0.7) {$\Omega R$};
%		\node at (-1.5,7.3) {$|\mathcal{A}|^2$};
%		\node at (6,13) {\small {$R \mathcal{M}_\phi = 0.008$}};
%		\end{tikzpicture}
%		\caption{The amplitude of outgoing radiation in the case $\Omega> \mathcal{M}$ for $\bar{M} = 0.008$. For $R=10\text{km}$ the horizontal axis shows a period in the range with $r_s/R=0.3$.}
%		\label{Amplitude}
%	\end{center}
%\end{figure}

\subsection{Approximate analytic solutions: near field, flat space limit}\label{Analytic}
We can also derive approximate analytic solutions for the axion profile by neglecting gravitational effects ($r_s=0$) and restricting to the near field limit (\ref{near}) corresponding to $r \Omega \ll 1$. The axion equation then becomes 
\begin{equation}
- \frac{d^2 \psi}{d r^2} + \left( \frac{\ell(\ell+1)}{r^2} + k^2 \right) \psi = \frac{R^2}{r^4}\theta(r -R). \label{analytic}
\end{equation}
This corresponds to the limit in which $r_s/R$ is significantly less than 1. Since pulsars have masses $M \sim M_\odot$ and $R=10\text{km}$, one typically has $r_s/R \sim 0.3$, meaning that the analytic solution we now describe gives a good order of magnitude estimate to the true result. The solution to this equation is determined by demanding continuity of $\psi$ and $\psi'$ at the stellar surface and imposing the boundary condition of regularity at the origin and decaying solutions or outgoing waves at infinity. For the specific case of $\ell=1$ considered here, this solution takes the form
\begin{equation}
\psi(r) =  \theta(R - r)\left[\frac{\kappa_1}{k r}\left(\sinh k r-k r\cosh k r \right)\right]+  \theta( r -R) \left[  \frac{\bar{\psi}(r)}{8}\frac{R^2}{r^2} +   \kappa_2 \, e^{-k r  } \left(1 +\frac{1}{r k}\right) \right]\label{ulytic},
\end{equation}
with 
\begin{align}
\bar{\psi} & =  kr \left[\text{Chi}(k  r ) \left[r  k  \cosh r  k -\sinh r  k \right]-  \text{Shi}(k  r ) \left[r  k 
	\sinh r  k -\cosh r  k \right]\right] -2,
\end{align}
where
\begin{align}
\text{Chi}(z)= \gamma_E +\log z+ \int_0^z dt \frac{\cosh t - 1}{t}, \qquad \qquad
\text{Shi}(z)= \int_0^z dt \frac{\sinh t}{t},
\end{align}
where $\gamma_E$ is Euler's constant and $\kappa_{1,2}$ are constants determined by demanding continuity of $u$ and $u'$ at the stellar surface given by
\begin{align}
\kappa_1 &= \frac{1}{8\bar{k}^2} \left(\bar{k}^4\text{Shi}(\bar{k} )-\bar{k}^4\text{Chi}(\bar{k} )+e^{-\bar{k} } \left(\bar{k}  \left(\bar{k} - 2 - \bar{k}^2\right)-2\right)\right),\nonumber \\
\kappa_2 &= \frac{1}{8 \bar{k} ^2}\Bigg(\left(\bar{k} ^2-2\right) \sinh \bar{k} +\left(\bar{k} ^2+2\right) \bar{k}  \cosh \bar{k} -\bar{k} ^4 \text{Shi}(\bar{k})\Bigg), \nonumber \\
\bar{k}& = R k . 
\end{align}
Plots of the different solution types are shown in fig.~\ref{AerialWave} together with a comparison with the approximate analytic result (\ref{ulytic}).

%\subsection{Resonant/Quasinormal modes}
%From the plots in figure \ref{umax} we see that as $\Omega \rightarrow \mathcal{M}_\phi$ the size of the peak of the $u_1$ solution reaches a maximum. This corresponds to a resonant solution given given by taking $k \rightarrow 0$
%\begin{equation}
%u_1^{\text{res}}(\rho) = \frac{(3-4 \rho ) \theta (\rho -1)-\rho ^4 \theta (1-\rho )}{12 \rho ^2}.
%\end{equation}
%The mangitude of the peak is given by $\psi_{11}^{rex}(\rho_{\text{max}}) = -1/9$, and occurs at $\rho_\text{max} = 3/2$.  

\section{Energy density and total mass}\label{MassSec}
The energy density of the scalar field is given by
\begin{align}
T_{00} & = \frac{1}{2} \dot{\phi}^2 + \frac{1}{2} \nabla\phi^2 + \frac{1}{2}\mathcal{M}_\phi^2 \phi^2,
\end{align}
where $T_{\mu \nu}$ is the energy momentum tensor for $\phi$. The largest values of the energy density correspond to the scenarios in which $\psi$ is greatest. This occurs for  $R \mathcal{M}_\phi \ll R \Omega \lesssim 1$ as shown in fig \ref{Peak}. In this limit, $\dot{\phi}^2 \sim \Omega^2 \phi^2$ and $\mathcal{M}^2 \phi^2$ are much less than $(\nabla \phi)^2 \sim R^{-2} \phi^2 $, in which case the energy density is dominated by the gradient term, so that, from (\ref{PhiAnsatz}) 
\begin{equation}
T_{00} \sim \frac{B^4_* g_{\gamma\phi}^2 R^6 \Omega^2}{r^2} \mathcal{O}\left(\psi^2\right)\mathcal{O}\left(Y_{\ell m}^2\right) \label{T00}.
\end{equation}
Thus near the surface of the star $r \approx R$, $\mathcal{O}(\psi^2)\approx 10^{-2}$ (see fig \ref{Peak}), so that the largest axion densities are characterised by
\begin{equation}
\rho_\text{axion} = 9.5 \times 10^{-2} \,\text{g} \, \,\text{cm}^{-3} \Bigg[\frac{B_*}{10^{13}\text{G}} \Bigg]^4\cdot
\Bigg[\frac{g_{\gamma \phi}}{10^{-13} \text{GeV}} \Bigg]^2\cdot
\Bigg[\frac{\Omega}{100 \text{Hz}} \Bigg]^2\cdot
\Bigg[\frac{R}{10 \text{km}} \Bigg]^4 \label{rhoAxion}.
\end{equation}
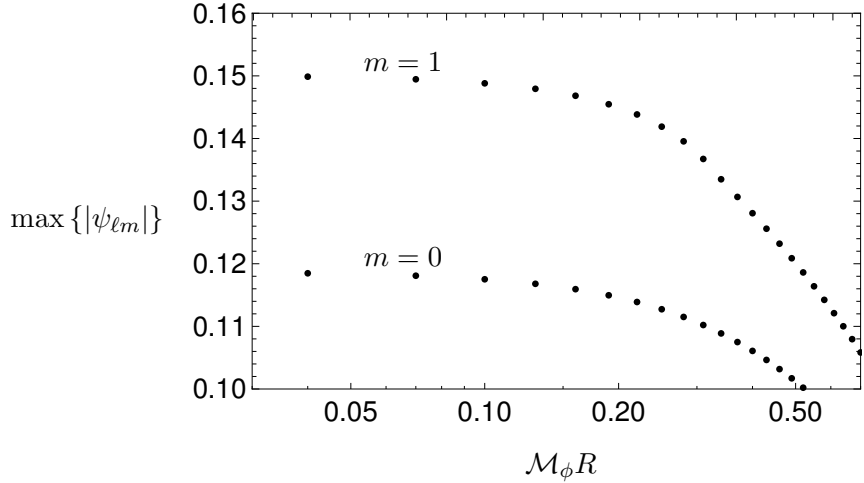
\begin{figure}[ht]
	\pgfdeclareimage[interpolate=true]{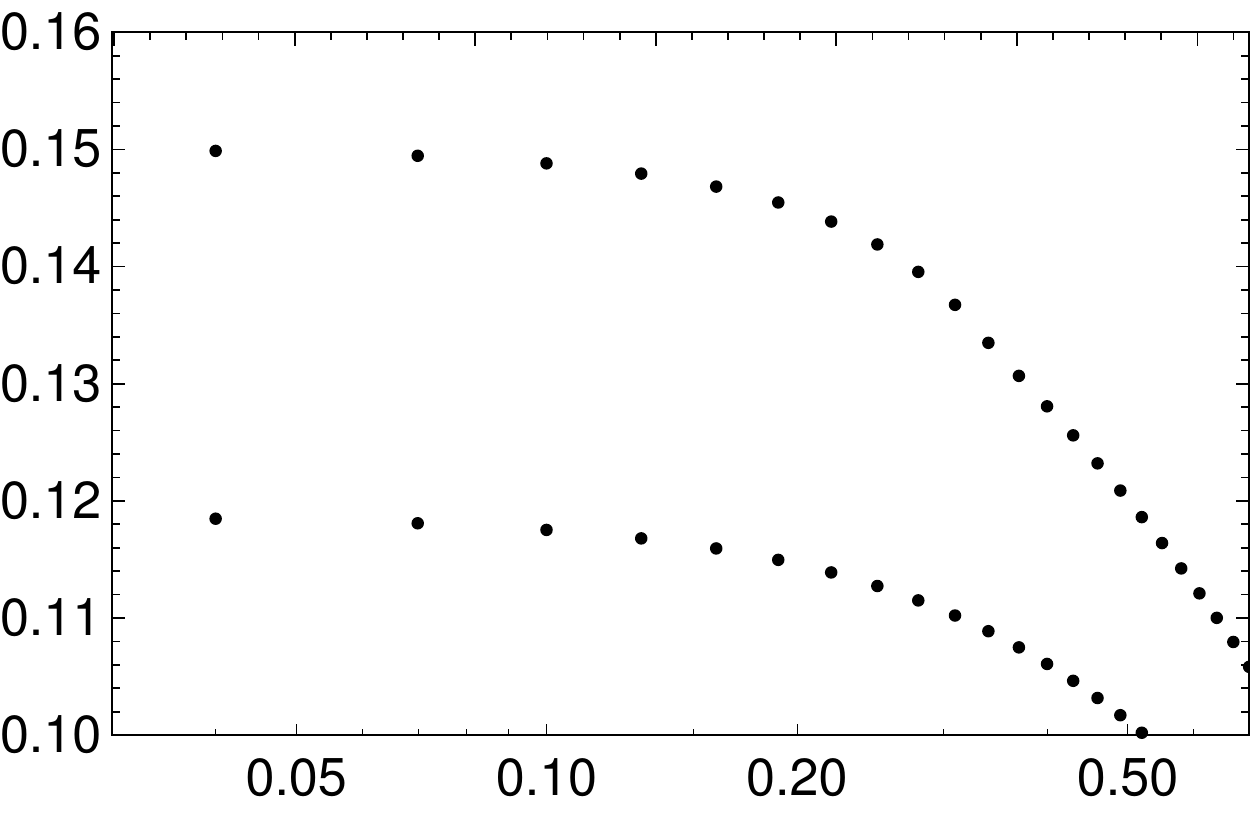}{PeakPlot.pdf}
	\begin{center}
		\begin{tikzpicture}[scale=0.7]
		\pgftext[at=\pgfpoint{0cm}{0cm},left,base]{\pgfuseimage{PeakPlot.pdf}}
		\node at (7,-0.7) {$\mathcal{M}_\phi R$};
		\node at (-2,4) {$\text{max}\left\{\left|\psi_{\ell m}\right|\right\}$};
		\node at (4,3.3) {$m =0$};
		\node at (4,7) {$m =1$};
		\end{tikzpicture}
		\caption{Maximum value of the radial function $|\psi_{\ell m}|$ as a function of $\mathcal{M}_\phi$ for fixed frequency $\Omega R =0.33$ and $r_s/R=0.3$}
		\label{Peak}
	\end{center}
\end{figure}

It is easy to see then that for typical pulsars, this can be considerably more than the ambient halo density of $10^{-24}\text{g}\text{cm}^{-3}$ \cite{Sikivie:1983ip}, even when $g_{\gamma \phi}$ is considerably below the exclusion boundary of roughly $10^{-13} \text{GeV}^{-1}$.

\begin{figure}[ht]
	\pgfdeclareimage[interpolate=true]{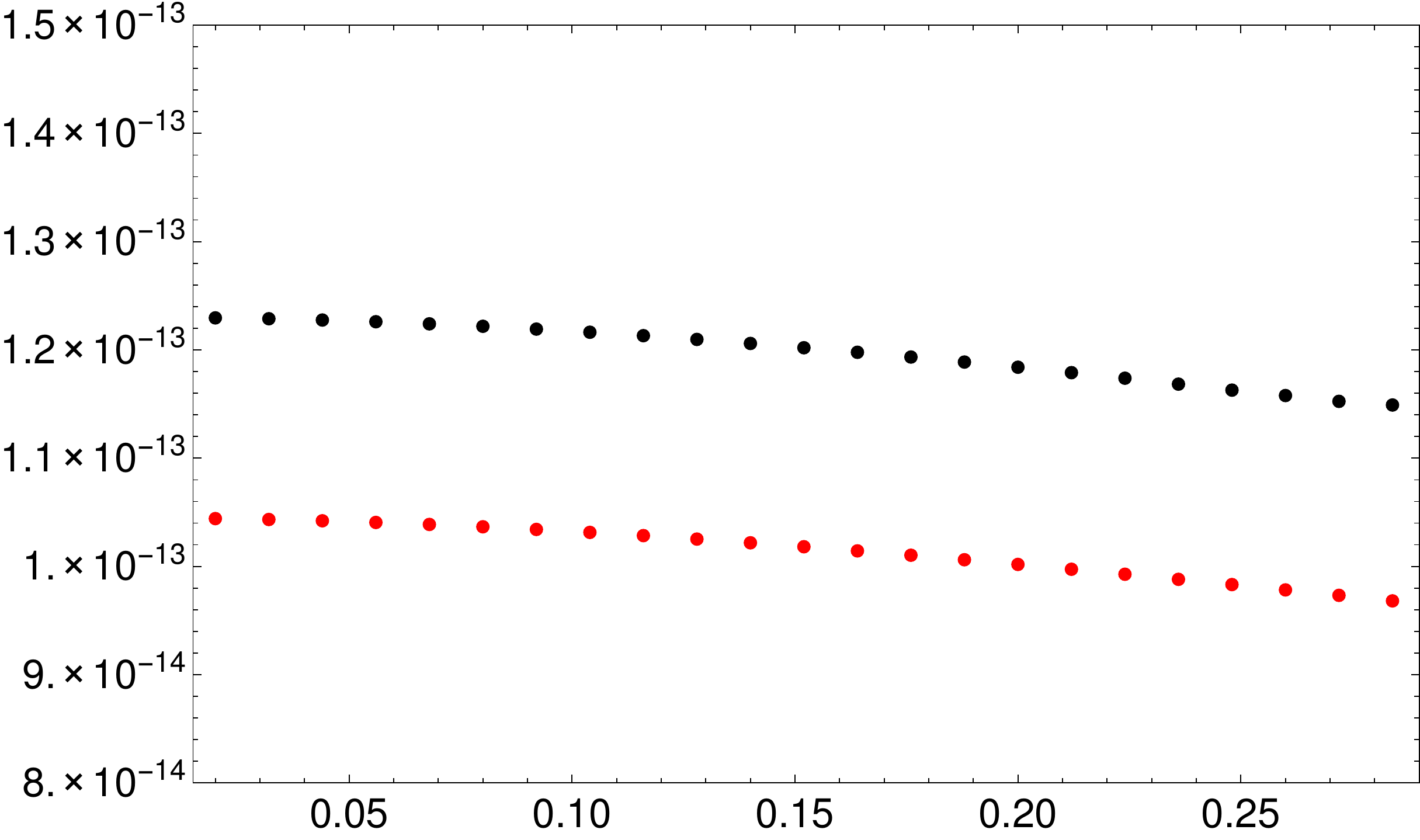}{MassPlot.pdf}
	\begin{center}
		\begin{tikzpicture}[scale=0.4]
		\pgftext[at=\pgfpoint{0cm}{0cm},left,base]{\pgfuseimage{MassPlot.pdf}}
%		\node[rotate=-90] at (19.4,10.8) {$\Omega = \mathcal{M}_\phi$} ;
		\node at (14,-0.9) {$ R \mathcal{M}_\phi$};
		\node at (-2,8) {$M/M_{_\odot}$};
		\end{tikzpicture}
		\caption{The total mass of the cloud in the maximal case for the values $B_*=10^{15} \text{G}$,  $g_{\gamma \phi}=10^{-13}\text{GeV}^{-1}$ and $R=20\text{km}$ with $\alpha = 45^{\circ}$ with a period $P=10\text{s}$. We also took $r_s/R=0.3$. The black dots show the full gravitational result, including radiative far-field corrections, and the red dots correspond to the analytic flat-space, near-field approximation. }
		\label{MassPlot}
	\end{center}
\end{figure}
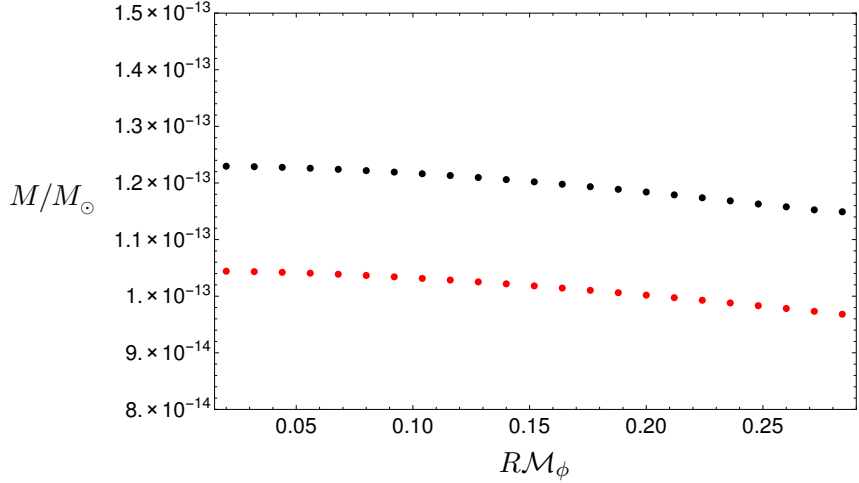

The energy density, $\delta \rho(r)$, contained in a spherical shell lying between $r$  and $r+ dr$ is given by
\begin{equation}
\delta\rho(r) = dr r^2 \int d\Omega T_{00}. \label{deltaRho}
\end{equation}
By inserting the general form (\ref{PhiAnsatz}) into (\ref{T00}) and (\ref{deltaRho}) a lengthy but straightforward calculation of the angular integral gives 
\begin{equation}
\frac{\delta \rho(r)}{r^2 \, dr }  =  
\frac{ B_*^4 g_{\gamma \phi}^2 \pi R^8 \Omega^2}{6} 
\sin^2 \alpha 
\Big[
\sin^2 \alpha \,F_{0}[\psi_{10}] + \cos^2 \alpha \left( F_{1}[\psi_{11,R}] + F_{1}[\psi_{11,I}] \right)
\Big],
\end{equation}
where
\begin{align}
F_m[\psi] & =\frac{1}{r^4} \left[ \psi^2 \left(3 + r^2 (\mathcal{M}_\phi^2 + m^2 \Omega^2) \right) - 2 r \psi \psi' + r^2 \psi'^2 \right],
\end{align}
where $\psi_{11, I}$ and $\psi_{11,R}$ denote the imaginary and real component of $\psi_{11}$ respectively. One can then define the total mass
\begin{equation}
M =  
\frac{ B_*^4 g_{\gamma \phi}^2 \pi R^8 \Omega^2}{24} 
\sin^2 \alpha 
\Big[
\sin^2 \alpha \,\int_0^\infty d r ^2 F_{0}[\psi_{10}] + \cos^2 \alpha \int_0^\infty d r r^2\left( F_{1}[\psi_{11,R}] + F_{1}[\psi_{11,I}] \right)
\Big].
\end{equation}
The total mass of the axion cloud is plotted in figure \ref{MassPlot} for the maximal case of a magnetar and the upper bound on the axion photon coupling. 

It is interesting to compare this to other astrophysical objects composed of axions, for instance, dilute axion stars examined in \cite{Braaten:2015eeu} have a range $M/M_{\odot} \leq 10^{-13}$,  and axion clusters have similar masses at around $M/M_{\odot} \sim 10^{-13}$ \cite{Enander:2017ogx}. What is of particular interest is the density of our axion configurations in comparison to axion stars in \cite{Braaten:2015eeu}. Our densities are of order $\rho \sim 10^{-1} \text{g}\text{cm}^{-3}$, for typical pulsars, slightly more dense than dilute axion stars shown in fig.~1 of ref. \cite{Braaten:2015eeu} which have $\rho \lesssim 10^{-2} \text{g}\text{cm}^{-3}$. The main advantage we have is that whilst existence of axion stars remains open, configurations around pulsars are a simple consequence of the axion-photon coupling. In this sense, they will inevitably be generated in regions around the pulsar in which $\textbf{E}\cdot \textbf{B}$ is sizeable. Furthermore, one has an axion cloud in the vicinity of strong magnetic fields, which could have interesting observational consequences \cite{Sikivie:1983ip} in terms of decays and stimulated decays via the axion-photon vertex.

This comparatively high axion density could have important implications in terms of observational signatures. However it is worth bearing in mind that a quantitative analysis will require a better understanding of the $\textbf{E}\cdot \textbf{B}$ profile under the influence of plasma screening in the magnetosphere of the pulsar. 

\section{Axion power output}\label{PowerSec}
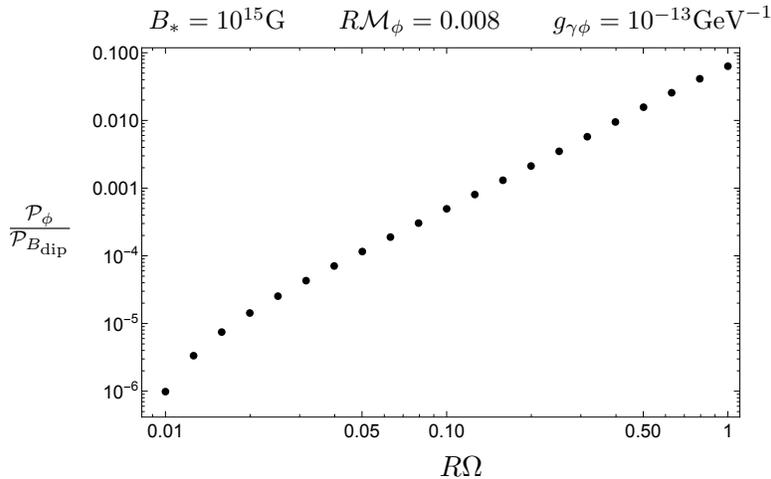
\begin{figure}[ht]
	\pgfdeclareimage[interpolate=true]{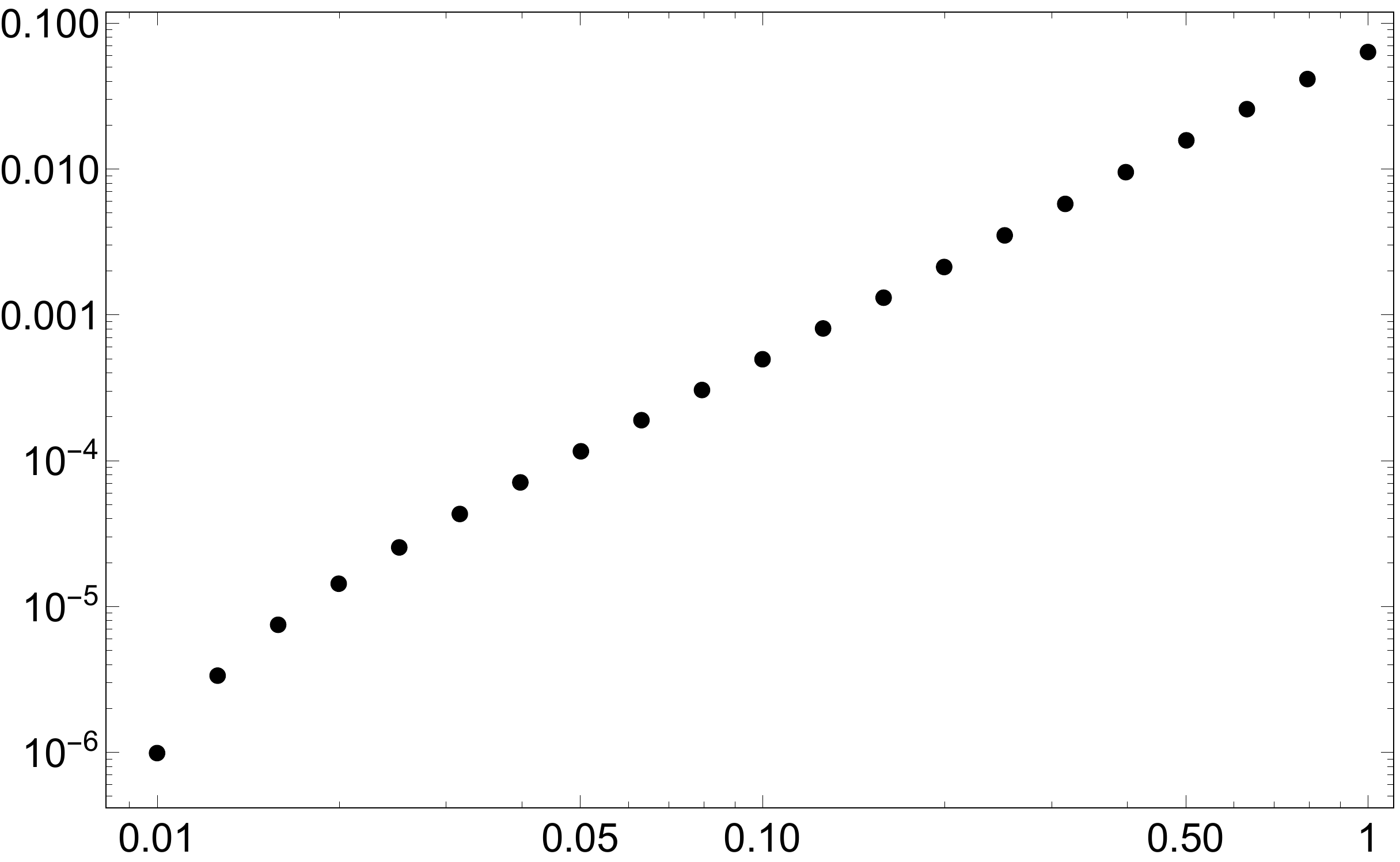}{AmpPlotRatio.pdf}
	\begin{center}
		\begin{tikzpicture}[scale=0.35]
		\pgftext[at=\pgfpoint{0cm}{0cm},left,base]{\pgfuseimage{AmpPlotRatio.pdf}}
		%		\node[rotate=-90] at (19.4,10.8) {$\Omega = \mathcal{M}_\phi$} ;
		\node at (14,-0.9) {$ R \Omega$};
		\node at (14,16) { \small $ B_* = 10^{15}\text{G}  \qquad R \mathcal{M}_\phi = 0.008  \qquad g_{\gamma \phi} = 10^{-13}\text{GeV}^{-1}$};
		\node at (-2,8) {$\frac{\mathcal{P}_\phi}{\mathcal{P}_{B_\text{dip}}}$};
		\end{tikzpicture}
	\end{center}
\caption{The ratio of electromagnetic dipole radiation to axion radiation as a function of $R \Omega$ for $B_* =10^{15}$G, $R\mathcal{M}_\phi = 0.008$, $g_{\gamma \phi} = 10^{-13}\text{GeV}^{-1}$ and $R=10\text{km}$.}
\label{Power}
\end{figure}

A further interesting question is how much power the axion configuration is extracting from the pulsar. This can be determined by computing the flux of the momentum density $\Pi_i \equiv T_{0i}$ through a sphere at radius $r$. The momentum density of the axion field is given by
\begin{equation}
\Pi = \dot{\phi} \nabla \phi.
\end{equation}
The energy flux per unit area across a sphere of radius $r$ is therefore
\begin{align}
f & = \frac{1}{4 \pi r^2 }\int_{S^2(r)} d \textbf{S} \cdot \nabla \phi \, \dot{\phi}   = \frac{1}{4 \pi }\int d\Omega_2 \,  \dot{\phi} \frac{\partial \phi}{\partial r}.
\end{align}
Note that $f$ has dimensions $[f] = \text{Energy} \cdot \text{Area}^{-1} \cdot \text{time}^{-1}$ and is different from the axion \textit{number} flux. The energy flux per solid angle is therefore given by
\begin{equation}
\frac{d f(r, \theta, \lambda)}{d \Omega_2} = \frac{1}{4\pi} \dot{\phi} \frac{\partial \phi}{\partial r} = - \frac{\Omega }{4\pi} \frac{\partial \phi}{\partial \lambda} \frac{\partial \phi}{\partial r} \label{dfdOmega}.
\end{equation}
Inserting the solution (\ref{PhiAnsatz}), 
%which can be written more explicitly as
%\begin{align}
%&\phi(r, \theta, \lambda) = \frac{g_{\gamma \phi}R^4 B_*^2\Omega}{4 r}\left[\sin^2 \alpha \,   \cos \theta \,\psi_{10} + \frac{\sin 2\alpha \sin \theta}{2}\left[\psi_{_{11,R}}   \sin \lambda  
%+\psi_{_{11,I}} \cos \lambda \right]\right],
%\label{overallSol}
%\end{align}
%with
%\begin{equation}
%\sin \lambda \sin \theta =  2 \sqrt{\frac{2 \pi}{3}} \text{Re}\left[ i Y_{11}\right], 
%\qquad
%\cos \lambda \sin \theta =  -2 \sqrt{\frac{2 \pi}{3}} \text{Re}\left[ Y_{11} \right]
%\end{equation}
the energy flux (\ref{dfdOmega}) averaged over the pulsar period is therefore given by
\begin{align}
\frac{d\braket{f}}{d\Omega_2} 
% =\frac{\Omega}{2\pi}\int_0^{2\pi/\Omega} d t \left[  - \frac{\Omega}{4\pi}  \frac{\partial \phi}{\partial \lambda} \frac{\partial \phi}{\partial r}\right]\nonumber  \\
 = \frac{1}{4 \pi r^2}  \frac{g_{\gamma \phi}^2 R^8 B_*^4 \Omega ^3}{128}  \sin^2 2\alpha  \sin^2 \theta \,\text{Im} \left[ \psi' \psi^* \right].
\end{align}
Notice therefore that there is only radiation if $\psi_{11}$ is complex, which requires $\Omega > \mathcal{M}$. In this case, from eq.~(\ref{uInfinity}) one has
\begin{equation}
\lim_{r\rightarrow \infty}  \text{Im} \left[ \psi' \psi^* \right]  = \left|\mathcal{A}( \Omega, \mathcal{M}_\phi) \right|^2 \sqrt{\Omega^2 - \mathcal{M}_\phi^2},
\end{equation}
which is constant, so that the radiation given by (\ref{dfdOmega}) falls off as $1/r^2$, as required by energy conservation. The power radiated by the axion cloud is therefore given by integrating this quantity over a sphere at infinity, which gives
\begin{equation}
\mathcal{P}_\phi =  \frac{1}{3}  \frac{g_{\gamma \phi}^2 R^8 B_*^4 \Omega^4}{64}  \sin^2 2\alpha  \left|\mathcal{A}( \Omega, \mathcal{M}_\phi) \right|^2 \sqrt{1 - \frac{\mathcal{M}_\phi^2}{\Omega^2}}.
\end{equation}
We now contrast this with the electromagnetic dipole radiation given by integrating the Poynting vector $\textbf{S}= \textbf{E} \times \textbf{B}$ (recalling we set $\mu_0=1$) over the sphere at infinity, which gives the power in dipole radiation
\begin{equation}
\mathcal{P}_{B_\text{dip}} = \frac{2 \pi}{3}   B_*^2 R^6 \Omega ^4 \sin ^2 \alpha  
\end{equation}
From this, it follows that 
\begin{align}
\frac{\mathcal{P}_\phi}{\mathcal{P}_{B_\text{dip}}} 
& = \frac{R^2 g_{\gamma\phi}^2 B^2_*}{32 \pi} \cos^2 \alpha  \left|\mathcal{A}( \Omega, \mathcal{M}_\phi) \right|^2 \sqrt{1- \frac{\mathcal{M}_\phi^2}{\Omega^2} } \nonumber \\
&= 9.6 \times 10^{-1} \times \Bigg[\frac{B_*}{10^{14}\text{G}} \Bigg]^2
\Bigg[\frac{g_{\gamma \phi}}{10^{-13} \text{GeV}} \Bigg]^2 
\Bigg[\frac{R}{10 \text{km}} \Bigg]^2 \cos^2\alpha \,\left|\mathcal{A}( \Omega, \mathcal{M}_\phi) \right|^2 
\sqrt{1- \frac{\mathcal{M}_\phi^2}{\Omega^2} }. \label{axionpower}
\end{align}
This ratio is plotted in fig.~\ref{Power}. In principle, since the derivative $\dot{\Omega}$ is determined by dissipation processes, as the pulsar loses rotational kinetic energy, one could use pulsar spindown measurements to place bounds on axion parameters. However in order to make reliable quantitative statements in this direction a better modelling of the screening of $\textbf{E}\cdot \textbf{B}$ in the pulsar magnetosphere will be needed. 

\section{Magnetospheric screening of $\textbf{E}\cdot \textbf{B}$} \label{screening}

The vacuum dipole model \cite{1955AnAp...18....1D} considered in previous sections assumes there is no plasma surrounding the neutron star, and that the electromagnetic fields satisfy the vacuum Maxwell equations (neglecting axion backreaction). However, this represents only a first approximation to a realistic pulsar model. Firstly, the VDM has only a trivial emission spectrum - there is only monochromatic dipole radiation with frequency identical to that of the pulsar and this ultra-low frequency radiation cannot be observed directly. That said, dissipation of rotational kinetic energy via dipolar radiation does play a role pulsar spin down \cite{2012hpa..book.....L}. 

In reality, observations show pulsars have a rich emission spectrum in observable frequency ranges arising from the acceleration of charged particles in their magnetic fields. Thus a realistic model of pulsars must take into account the collection of surrounding charges, currents and their corresponding electromagnetic fields, which together form the pulsar \textit{magnetosphere}. Recent reviews of pulsar magnetospheres can be found in \cite{Melrose:2016kaf,Petri:2016tqe,Cerutti:2016ttn}, and an excellent and condensed history of the key developmental stages is given by \cite{Harding:2017ttz}.

Even in the early days of pulsar theory it was pointed out by Goldreich and Julian \cite{Goldreich:1969sb} that the electromagnetic forces on particles at the stellar surface exceed that of gravity by many orders of magnitude. As a result, particles will be inevitably lifted from the stellar surface and into the surrounding atmosphere. Early work \cite{Goldreich:1969sb} treated the magnetosphere as perfect conductor in the following sense. As explained at the beginning of sec.~\ref{PulsarB} a perfectly conducting plasma must satisfy
\begin{equation}
 \textbf{E} + \textbf{v} \times \textbf{B} = 0. \label{Conductor}
\end{equation}
As a result, particles with charge $q$, corotating with the pulsar have velocity $\textbf{v}$ and experience no Lorentz force
\begin{equation}
\textbf{F} = q \left( \textbf{E} + \textbf{v} \times \textbf{B} \right) =0.
\end{equation}
Furthermore, one sees immediately that $\textbf{E} \cdot \textbf{B} =0$ so that there is no component of force in the direction of $\textbf{B}$, and particles slide freely along magnetic field lines. The electric field satisfying (\ref{Conductor}) has a corresponding charge density given by Gauss' law \cite{Melrose:2011eh}
\begin{equation}
	\rho = \nabla \cdot \textbf{E} = -2 \boldsymbol{\Omega}\cdot \textbf{B} + \textbf{v} \times \textbf{B}. \label{density}
\end{equation}
Notice that we have not specified anything about the form of the magnetic field, so that eq.~(\ref{density}) remains entirely general. If one specialises to a magnetic dipole, one obtains the classic Goldreich Julian density \cite{Goldreich:1969sb}. The interpretation is that this distribution corresponds to charges arranging themselves in the plasma in such a way as to exactly screen the component $\textbf{E}\cdot \textbf{B}/B$ of the induced electric field arising from the time-dependent $\textbf{B}$ field.

If it were true that everywhere in the pulsar magnetosphere $\textbf{E}\cdot \textbf{B}$ vanishes, two things would follow. Firstly, there would be no acceleration of particles along magnetic field lines and secondly, the source term for the axion field (\ref{AEOM}) would vanish. However, in order to explain observed emission spectra of pulsars, one must have regions in which particles \textit{do} experience such acceleration, producing radiation in the process. Regions in which $\textbf{E}\cdot\textbf{B}$ is non-vanishing are known variously as acceleration regions, or acceleration gaps, with the term ``gap" arising from the fact the plasma density may deviate from (\ref{density}). 

Understanding the exact location of acceleration gaps, the size of $\textbf{E}\cdot \textbf{B}$ within them, and the acceleration of charges in these regions has been a major effort in pulsar magnetosphere modelling in recent decades. One approach is to relax the constraint of perfect conductivity embodied in the condition (\ref{Conductor}). This can be achieved by incorporating a finite conductivity $\sigma$ into the electrodynamic equations. This is known as resistive pulsar magnetosphere modelling, see e.g.~refs.~\cite{Li:2011zh, Kalapotharakos:2011vg}. Such a setup allows one to formulate a semi-analytic approach (the equations of motion must still be solved numerically). Although, resistive pulsar magnetosphere models are able to reproduce observed properties of pulsar emission spectra \cite{Kalapotharakos:2013sma,Brambilla:2015vta}, their drawback is that the exact form of the conductivity has essentially to be added by hand.

 It therefore becomes necessary to treat the pulsar magnetosphere and its constituent plasma in a fully self-consistent way, by solving the full set of electrodynamic equations for a general current and EM fields. In addition one should allow for the injection of particles from the stellar surface and pair creation in the magnetosphere. Recent success in this direction utilises the ``particle-in-cell" approach \cite{Philippov:2013tpa,Chen:2014dva,Chen:2016qkk,Philippov:2014mqa,Cerutti:2014ysa,Kalapotharakos:2017bpx, Brambilla:2017puc,Levinson:2018arx} which enables one to determine in a more precise way the location of the acceleration gaps and regions in which $\textbf{E}\cdot\textbf{B} \neq 0$.

In light of this, our paper presents some of the key methodology needed to model axion clouds around pulsars and identifies some of the main qualitative features, e.g. the radiative and bound state solutions, and their scaling with axion mass and pulsar frequency. Generally speaking, we would expect the size and spatial extent of $\textbf{E}\cdot \textbf{B}$ (and therefore our effect) to be strongest for pulsars with low plasma densities, but higher magnetic fields, such as young pulsars - see figures 9 and 13 and table 2 of ref. \cite{Kalapotharakos:2017bpx}.

However, in moving towards a more realistic pulsar model and its corresponding axion clouds, one should precisely identify the acceleration gaps with $\textbf{E}\cdot \textbf{B} \neq 0$. One possibility, as a first step, would be to produce the $\textbf{E}\cdot \textbf{B}$ profile in one of the resistive pulsar magnetosphere models, decompose into spherical harmonics as in eq.~(\ref{YDecomp}) and extract the corresponding axion solutions. One could then (at least in the linear regime) determine the axion profile on a mode-by-mode basis. Such an approach has the advantage that it is a straightforward generalisation of the present calculation and also allows for a simple semi-analytic treatment of the electrodynamics with finite conductivity. Eventually this could in principle be extended to a full particle-in-cell simulation, though we shall leave such considerations for future work.

\section{Summary and discussion and possible observable effects}\label{discsusion}

In this paper we examined the size of axion solutions sourced by the axion-photon coupling in the electromagnetic fields of neutron stars, using a simple vacuum dipole model. To determine the axion profile, we solved the inhomogeneous wave equation with a source term given by $\textbf{E}\cdot \textbf{B}$. We found that in the extremal case, for the largest known magnetic fields, achieved by magnetars, $B_* = 10^{15}\text{G}$ and an axion-photon coupling of $g_{\gamma \phi} \sim 10^{-13} \text{GeV}^{-1}$, that the total mass of the cloud can be of order $10^{-13}M_\odot$, and density reaching $\rho \sim 10^{-1} \text{g}\,\text{cm}^{-3}$, which compares favourably to certain classes of axion stars. We  computed the total power output in the radiative case and speculated on how this might be used to constrain axion physics if sufficiently accurate measurements of pulsar spin down can be made.  

This paper therefore makes the first important steps towards mapping out the structure of axion clouds around pulsars sourced by the axion coupling to electromagnetism. As alluded to in sec.~\ref{screening} the next step is to incorporate magnetospheric screening of $\textbf{E}\cdot \textbf{B}$ to understand more precisely in what regions around the pulsar, and for what classes of pulsar $\textbf{E}\cdot \textbf{B}$, is most sizeable. This will allow a better quantitative understanding of the effects presented here allowing for a more accurate discussion of observational signatures. There are a number of possible effects which should be investigated which we mention below. 
 
\begin{itemize}
	\item \textbf{Polarisation and birefringence.} Since the axion-photon coupling violates parity, photons passing through a medium of axions experience different dispersions relations depending on whether they are left or right polarised. Thus axion clouds behave as a birefringent medium. This has already been examined for light passing through axion clusters \cite{Tkachev:2014dpa} and also through a superradiant cloud around a black hole \cite{Plascencia:2017kca}. There are a number of possibilities which should be pursued in the present case. Firstly, it is in precisely the ``acceleration regions" where $\textbf{E}\cdot\textbf{B} \neq 0$  that pulsar emission spectra are generated due to accelerated charges. Therefore the presence of non-vanishing axion configurations in these regions may have polarisation effects on pulsar emissions. Similarly, in the same vein as \cite{Plascencia:2017kca}, one could look at light from a known linearly polarised source passing close by the pulsar and experiencing birefringence. One final possibility is that in the $\Omega > \mathcal{M}$ regime where outgoing axion wave fronts are produced, one could examine how photons emitted from the pulsar ``ride" the outgoing axion waves, experiencing polarisation effects on their journey from the pulsar. This would be an integrated effect over the light's journey from the pulsar to a detector.
	\item \textbf{Luminosity and spontaneous decay of axions.} It was noted some time ago that axion clouds can glow via the decays of axions into two photons with frequency $\mathcal{M}_\phi/2$. This was exploited in the case of axion clusters \cite{Kephart:1986vc,Kephart:1994uy} and more recently for superradiant axion clouds around black holes \cite{Rosa:2017ury}. However, one would need an axion cloud of significant mass to produce an observable luminosity and also a high axion mass to give a large decay rate which goes as $\Gamma_{\phi \rightarrow \gamma \gamma}= \mathcal{M}_\phi^3 g_{\gamma \phi}^2/64\pi$. In our case, large axion masses give highly suppressed cloud sizes, whilst smaller masses give a tiny decay width, thus it seems unlikely that the luminosity will be significant. One also would need to understand from a theoretical point of view, how dissipation into photons can be incorporated into the field equations as quantum corrections. Presumably one would expect the imaginary part of the one-loop photon correction to the axion propagator to give rise to a dissipative term as in e.g. \cite{Yokoyama:2005dv}.
	
	\item \textbf{Stimulated Decays.} One major advantage of studying axion configurations in the vicinity of pulsars is that they possess tremendously strong electromagnetic fields. This could have interesting consequences in terms of stimulated axion to photon conversions in the presence of the classical electromagnetic fields. The signatures arising from such processes in the vicinity of neutron stars have been examined in \cite{Pshirkov:2007st,Huang:2018lxq,Fortin:2018ehg,2018arXiv180403145H}. This exploits the same physics as light shining through a wall and haloscopes experiments \cite{Sikivie:1983ip,Adler:2008gk}.

%	\item \textbf{Detection of In the $\Omega > \mathcal{M}_\Phi$ case, can we detect the outgoing axion wavefronts directly via GW detctors such as those proposed by Arvanitaki \cite{Arvanitaki:2016fyj}?
	\item \textbf{Generation of secondary electromagnetic fields}. The axion solution can back-react onto electromagnetic fields via their equations of motion $\partial_\mu F^{\mu \nu} = -g_{\gamma \phi} \partial_\mu \phi \tilde{F}^{\mu \nu}$. Thus the primary magnetic fields of the neutron star and the axion cloud form an effective current $j^\mu_\text{eff} \equiv - \partial_\nu \phi \tilde{F}^{\nu \mu}$ which can source secondary electromagnetic fields. For instance, in the present VDM model for the pulsar, both $\phi$ and $A_\mu$ contain a $Y_{1 1}$ harmonic, so the effective charge $\rho_\text{eff}  = - g_{\gamma \phi}\left(\textbf{B} \cdot \nabla{\phi} \right)$ will contain a corotating quadrupole moment $Y_{22}(\theta, \lambda)$, with $\lambda = \varphi - \Omega t$. This could give rise to a secondary sub-pulse from the pulsar in between the main beam. Firstly though, one would need a better pulsar model incorporating magnetospheric effects and, from an observational point of view, one should also take into account any higher multipole moments of  neutron stars  \cite{Petri:2015oaa} which may obscure any signal coming from axion back-reaction. 
%	\item Could the axion cloud make a lump in the NS surface via gravitational tidal forces, creating a bump in the neutron star surface which would show up in the form of continuous GWs?
\end{itemize}
Further effects which should be investigated are bosanova phenomena for high axion densities when non-linear self-interactions become important \cite{Yoshino:2012kn,Yoshino:2015nsa}. This can lead to supernova style explosions and bouncing solutions. Another interesting possibility is to study how our clouds modify the forces between neutron stars as in \cite{Hook:2017psm} and how our discussion might modify the nature of the solutions present therein. 

%It will also be interesting to examine dissipation effects in greater detail as has been done in the context of superradiance for stars \cite{Cardoso:2015zqa}.

\section*{Acknowledgements}
JIM would like to thank Carlos Tamarit, Juan Sebasti{\'a}n Cruz and Juraj Klari{\'c} for their help with numerics and discussions on axion physics, as well as Francesca Day for advice on axion constraints and Jochen Greiner for his comments on neutron stars. He also wishes to thank Gianmassimo Tasinato for first bringing axion physics in astrophysical settings to his attention and Francesco Filippini and Carlos Nunez for discussions during the early stages of this work. We also thank Hong Zhang for comments on the draft of this manuscript. We are also grateful to the Sonderforschungsbereichen (SFB) 1258 for providing a space for stimulating discussions. This work has benefited from the support of a Technical University Foundation Fellowship. 

\section*{Appendices}
\begin{appendices}
	
	\section{Green Functions}\label{Green}
	The Green functions for eq.~(\ref{Grstar}) can be expanded in terms of the two linearly independent mode functions $f(r)$ and $g(r)$ which we choose to have boundary conditions given by eq.~(\ref{Asymptotics}). The Green function therefore takes the form
	\begin{equation}
	G(r,r' ) =  
	\left \{
	\begin{tabular}{cc}
	$a(r') f(r) +b(r') g(r)$  & $r>r'$\\
	$c(r') f(r)  +d(r') g(r)$ & $r<r'$
	\end{tabular}
	\right. ,
	\end{equation}
	for some functions $a$,$b$,$c$ and $d$ depending on $r'$. 	The Green function must also satisfy the usual continuity and jump conditions given by
	\begin{align}
	&\lim_{r \rightarrow r^{\prime +}}G(r,r') = \lim_{r \rightarrow r^{\prime -}}G(r,r'), \\
	&\lim_{r \rightarrow r^{\prime +}}\frac{\partial}{\partial r}G(r,r') - \lim_{r \rightarrow r^{\prime -}}\frac{\partial}{\partial r}G(r,r') = - \frac{1}{A(r')^2}. \end{align}
	Imposing these constraints, one can eliminate $c$ and $d$ from the Green function to get
	\begin{equation}
	G(r,r') =  
	\left \{
	\begin{tabular}{cc}
	$a(r') f(r) +b(r') g(r)$  & $r>r'$\\
	$\left[a(r') - \frac{g(r')}{A(r')^2W(r')} \right]f(r)  + \left[b(r')+ \frac{f(r')}{A(r')^2W(r')} \right] g(r)$ & $r<r'$
	\label{GreenFunc}
	\end{tabular}
	\right. ,
	\end{equation}
	where $W\left(r\right)$ is the Wronskian given explicitly by
	\begin{equation}
	W\left[f(r),g(r)\right] \equiv f(r) g'(r)-f'(r)g(r). \label{Wronskian}
	\end{equation} 	
	Next, by setting $b(r') =0$ we impose the condition of decaying solutions or outgoing waves in the cases ($\mathcal{M} > \Omega$) and ($\mathcal{M} < \Omega$), respectively. The Green function then takes the form 
	\begin{equation}
	G(r,r' ) =  
	\left \{
	\begin{tabular}{cc}
	$a(r') f(r)$  & $r>r'$\\
	$\left[a(r') - \frac{g(r')}{A(r')^2W(r')} \right]f(r)  + \frac{f(r')}{A(r')^2 W(r')} g(r)$ & $r<r'$.
	\end{tabular}
	\right. .
	\end{equation}
	The source term and the Wronskian decay only as an inverse power of $r$, however $g(r)$ grows exponentially. It follows that when integrating over the region $r' > r$,  the $g(r')$ contribution gives a divergent integral. We therefore must cancel this behaviour from $g$ by taking
	\begin{equation}
	a(r') =  \gamma(r')+\frac{g(r')}{A(r')^2W(r')} ,
	\end{equation} 
	where $\gamma(r')$ is some function decaying suitably fast so as to make all integrals over $r'$ convergent. The Green function therefore takes the form
	\begin{equation}
	G(r_,r') =  
	\left \{
	\begin{array}{cc}
	\gamma(r')f(r) + \frac{g(r') f(r)}{A(r')^2W(r')}   & r >r'\\
	\gamma(r') f(r) + \frac{g(r)f(r')}{A(r')^2W(r')} &  r < r'
	\end{array}
	\right.
	. \label{Gform}
	\end{equation}
Next we recall that the differential operator 
\begin{equation}
L = -A\frac{d}{d r} \left( 
A\frac{d  }{d r}
\right) + V_\ell(r)  -  m^2 \Omega^2,
\end{equation}
appearing on the left hand side  of (\ref{Grstar}), is of the Sturm-Liouville type, so that it is self-adjoint with respect to following inner product with weight $A(r)^{-1}$, given by
\begin{equation}
\braket{\psi_1, \psi_2} = \int \frac{dr}{A(r)} \psi_1(r) \psi_2(r).
\end{equation}
In other words, $\braket{\psi_1, L \psi_2} = \braket{L \psi_1, \psi_2}$. It therefore follows that since $G(r,r')$ is the inverse of a self-adjoint operator that $G$ is also self-adjoint, since $\braket{G \psi_1, \psi_2} = \braket{G \psi_1, L G \psi_2} = \braket{\psi_1, G \psi_2}$. Since this holds for any $\psi_1$ and $\psi_2$, it therefore follows that $G(r,r')/A(r)$ is symmetric with respect to $r \leftrightarrow r'$. By this symmetry, it follows that $\gamma(r') = \gamma f(r')/A(r)$ for some constant $\gamma$. Furthermore, since $W(r)A(r) = 2k$, one sees that $G(r,r')$ now takes the form
\begin{equation}
G(r_,r') =   \frac{\gamma}{A(r')}  f(r) f(r')+ \frac{1}{2k A(r')}\left[
\theta(r - r')g(r') f(r)  + 
\theta(r' - r)g(r) f(r') \right],
\end{equation}
which manifestly has the desired symmetry properties, where the constant $\gamma$ is fixed by matching the exterior and interior solutions.

	\section{Interior Solution}\label{InteriorSol}
	The interior wave equation (\ref{uEqInterior})	 can be recast as
	 \begin{equation}
	 r^2 u'' +  r P(r) u' + Q(r) u = 0,
	 \label{EOMuINterior}
	 \end{equation}
	 where $P$ and $Q$ are given by eqs.~(\ref{PQeq}).
	 Notice that $P$ and $Q$ are analytic about $r =0$. It therefore follows that $r=0$ is a regular singular point since 
	 \begin{equation}
	 \lim_{r \rightarrow 0}  P(r) = 0, \qquad \lim_{r \rightarrow 0} Q(r)  = - \ell(\ell + 1) , \label{PQlimits}
	 \end{equation}
	 both give finite limits. We therefore use a Frobenius method and seek solutions of the form
	 \begin{equation}
	 \psi_\text{int}(r) = R^{-s} r^{s} \tilde{\psi}(r) \label{uinterioransatz},
	 \end{equation}
	 where $s$ satisfies the indicial equation
	 \begin{equation}
	 s(s-1) + P(0) s + Q(0) = 0,
	 \label{sroot}
	 \end{equation}
	 where $\tilde{\psi}(r)$ is analytic about $r=0$. From eq.~(\ref{PQlimits}) it is easy to see that the roots of the equation (\ref{sroot}) are given by
	 \begin{equation}
	 s_\pm = \frac{1}{2}\left[ 1 \pm \sqrt{1 + 4 \ell (\ell+1)} \right]. 
	 \end{equation}
 Inserting (\ref{uinterioransatz}) into (\ref{EOMuINterior}) gives an equation of motion for $\tilde{u}$ of the from:
	 \begin{equation}
	 r^2 \tilde{\psi}'' + r \left( P + 2 s\right) \tilde{\psi}'+ \left[ s(s-1) + s P + Q \right]\tilde{\psi} =0.\label{geq}
	 \end{equation}
	 The general solution in the stellar interior can then be expressed as a linear combination of solutions $\tilde{\psi}_\pm$ corresponding to these two roots. Note that for the specific case $\ell=1$ we consider here, the roots are integers with $s_+=2$ and $s_- = -1$ so that the full solution takes the form
	 \begin{equation}
	 \psi_\text{int}(r) = r^2  \tilde{\psi}_+(r) + r^{-1} \tilde{\psi}_-(r).
	 \label{uint}
	 \end{equation}
	 Expressing each $\tilde{\psi}$ as a power series
	 \begin{equation}
	 \tilde{\psi}(r) = \sum_{n=0}^\infty a_n r^n, \label{powerg}
	 \end{equation}
	 we identify $\tilde{\psi}(0) = a_0$ and $\tilde{\psi}'(0) = a_1$ and after substituting the ansatz (\ref{powerg}) into (\ref{geq}), we find the following relation between $a_0$ and $a_1$ of the form
	 \begin{equation}
	 \left(P(0) + 2 s\right) a_1 + \left(s P'(0) + Q'(0) \right) a_0 =0.
	 \end{equation}
	 This gives a relation between $\tilde{\psi}(0)$ and $\tilde{\psi}'(0)$: 
	 \begin{equation}
	 \tilde{\psi}'(0) = \frac{s P'(0) + Q'(0)}{P(0) + 2s } \tilde{\psi}(0).
	 \end{equation}
	 A short calculation shows that $P'(0)=Q'(0)=0$ and so $\tilde{\psi}'(0)=0$. If we impose the further constraint that $u$ is regular at the origin, the $s=-1$ contribution $\psi \sim r^{-1} \tilde{\psi}_-$, implies that $\tilde{\psi}_- (0) = 0$, which means that $\tilde{\psi}_-$ therefore vanishes everywhere in the stellar interior. This means that the interior solution must take the form
	 \begin{equation}
	 \psi_{\text{int}}(r) = r^2  \tilde{\psi}_+(r), \qquad 0 \leq r \leq R \label{uint}
	 \end{equation} 
	 where $\tilde{\psi}_+(r)$ satisfies eq. (\ref{geq}) with $s=2$ subject to the boundary condition $\tilde{\psi}_+'(0)=0$. 
\end{appendices}

\bibliography{References_Axion_Configurations}{}
\bibliographystyle{JHEP}

\end{document}